\documentclass[floats,floatfix,twocolumn,superscriptaddress,nofootinbib,showpacs]{revtex4-1}

\pdfoutput=1
\usepackage{amsfonts,amssymb,amsmath}
\usepackage[usenames,dvipsnames]{xcolor}
\usepackage[unicode,colorlinks=true,linkcolor=blue,urlcolor=blue,citecolor=gray]{hyperref}
\usepackage[all]{hypcap}
\usepackage{graphicx}
\usepackage{xspace}
\usepackage{tikz}
\usetikzlibrary{shapes,arrows}
\usetikzlibrary{matrix}
\usetikzlibrary{shapes.multipart}
\usetikzlibrary{er,positioning}
\usepackage[normalem]{ulem} 
\usepackage{url}
\usepackage{color,soul}
\usepackage[caption=false]{subfig}
\usepackage[T1]{fontenc}
\usepackage[utf8]{inputenc}
\usepackage{microtype}

\newcommand{\Caltech}{\affiliation{Theoretical Astrophysics,
    Walter Burke Institute for Theoretical Physics,\\
    California Institute of Technology, Pasadena, CA 91125, USA}}
\newcommand{\Cornell}{\affiliation{Center for Radiophysics and Space
    Research, Cornell University, Ithaca, New York 14853, USA}}
\newcommand{\OleMiss}{\affiliation{Department of Physics and Astronomy, The University of Mississippi, University, MS 38677, USA}}

\newcommand{\bs}[1]{\boldsymbol{#1}}

\newcommand{\eps}{\ensuremath{\varepsilon}}

\newcommand{\pd}{\partial}

\newcommand{\nn}{\nonumber}

\newcommand{\txt}[1]{{\textrm{\tiny{#1}}}}
\newcommand{\mpl}{\ensuremath{m_\txt{pl}}}

\newcommand{\dual}{\,{}^*\!}

\begin{document}

\title{Numerical binary black hole collisions in dynamical Chern-Simons gravity}

\author{Maria Okounkova}
\email{mokounko@tapir.caltech.edu}
\Caltech
\author{Leo C. Stein}
\OleMiss
\author{Mark A. Scheel}
\Caltech
\author{Saul A. Teukolsky}
\Caltech \Cornell

\date{\today}

\begin{abstract}
We produce the first numerical relativity binary black hole gravitational waveforms in a higher-curvature theory beyond general relativity. In particular, we study head-on collisions of binary black holes in order-reduced dynamical Chern-Simons gravity. This is a precursor to producing beyond-general-relativity waveforms for inspiraling binary black hole systems that are useful for gravitational wave detection. Head-on collisions are interesting in their own right, however, as they cleanly probe the quasi-normal mode spectrum of the final black hole. We thus compute the leading-order dynamical Chern-Simons modifications to the complex frequencies of the post-merger gravitational radiation. We consider equal-mass systems, with equal spins oriented along the axis of collision, resulting in remnant black holes with spin. We find modifications to the complex frequencies of the quasi-normal mode spectrum that behave as a power law with the spin of the remnant, and that are not degenerate with the frequencies associated with a Kerr black hole of any mass and spin. We discuss these results in the context of testing general relativity with gravitational wave observations.
\end{abstract}

\maketitle

\section{Introduction}
\label{sec:introduction}

At some length scale, Einstein's theory of general relativity (GR) must break down and be reconciled with quantum mechanics in a  beyond-GR theory of gravity.  Binary black hole (BBH) mergers probe the strong-field, non-linear regime of gravity, and gravitational waves from these systems could thus contain signatures of such a theory. Current and future gravitational wave detectors have the power to test GR~\cite{Berti:2015itd}, and BBH observations from LIGO and Virgo have given a roughly 96\% agreement with GR~\cite{TheLIGOScientific:2016src, Abbott:2017oio}. 

These tests of GR, however, are presently null-hypothesis and parametrized tests~\cite{Yunes:2016jcc, TheLIGOScientific:2016src}, which use gravitational waveforms produced in GR with numerical relativity. An open problem is the simulation of BBH systems through full inspiral, merger, and ringdown in beyond-GR theories. Waveform predictions from such simulations would allow us to perform \textit{model-dependent} tests, and to parametrize the behavior at merger in beyond-GR theories. 

In this study, we consider dynamical Chern-Simons (dCS) gravity, a beyond-GR effective field theory that adds a scalar field coupled to spacetime curvature to the Einstein-Hilbert action, and has origins in string theory, loop quantum gravity, and inflation~\cite{Alexander:2009tp, Green:1984sg, Taveras:2008yf, Mercuri:2009zt, Weinberg:2008hq}. Computing the evolution of a binary system requires first specifying suitable initial conditions. Because the well-posedness of the initial value problem in full dCS gravity is unknown~\cite{Delsate:2014hba}, we work instead in a well-posed \textit{order-reduction scheme}, in which we perturb the metric and scalar field around a GR background~\cite{Okounkova:2017yby}. The leading-order modification to the spacetime metric, and hence gravitational radiation, occurs at second order, which is precisely the order we consider in this study, building on our previous work~\cite{Okounkova:2017yby, MashaIDPaper, MashaEvPaper}.

While our ultimate goal is to produce full inspiral-merger-ringdown waveforms relevant for astrophysical BBH systems, in this study we consider the leading-order dCS corrections to binary black hole head-on collisions. Such configurations, while less astrophysically relevant than orbiting binaries, serve as a proof of principle for our method of producing BBH waveforms in a beyond-GR theory~\cite{MashaEvPaper}, and are fast and efficient to run. Head-on collisions also contain interesting science in their own right, as they cleanly probe the quasi-normal mode (QNM) spectrum of the post-merger gravitational radiation~\cite{Anninos:1993zj, Anninos:1995vf, Baker:2000zh, Sperhake:2005uf}. In this study, we thus produce the first BBH waveforms in a higher-curvature beyond-GR theory, and probe the leading-order dCS modification to the QNM spectrum of a head-on BBH collision.

\subsection{Roadmap and conventions} 
\label{sec:conventions}

This paper is organized as follows. We give an overview of our methods in Sec.~\ref{sec:methods}, 
and refer the reader to previous papers,~\cite{MashaEvPaper} and~\cite{MashaIDPaper}, as well as Appendices~\ref{sec:GaugeAppendix} and~\ref{sec:WaveExtractionAppendix}, for technical details. We discuss fitting perturbed quasi-normal modes in Sec.~\ref{sec:QNMsec}. We present and discuss our results, including quasi-normal mode fits, in Sec.~\ref{sec:results}. We discuss the implications of this  study on  testing GR in Sec.~\ref{sec:testingGr}. We conclude in Sec.~\ref{sec:conclusion}.

We set $G = c = 1$ throughout. Quantities are given in terms of units of $M$, the sum of the Christodoulou masses of the background black holes at a given relaxation time~\cite{Boyle:2009vi}. Latin letters in the beginning of the alphabet $\{a, b,  c, d \ldots \}$ denote 4-dimensional spacetime indices, while Latin letters in the middle of the alphabet $\{i, j ,k, l, \ldots \}$ denote 3-dimensional spatial indices (present in the appendices). $g_{ab}$ refers to the spacetime metric with connection $\Gamma^a{}_{bc}$, while $\gamma_{ij}$ (used in the appendices) refers to the spatial metric from a 3+1 decomposition with corresponding timelike unit normal one-form $n_a$ (cf.~\cite{baumgarteShapiroBook} for a review of the 3+1 ADM formalism). 


\section{Methods}
\label{sec:methods}

\subsection{Order-reduced dynamical Chern-Simons gravity}
\label{sec:dCS}

Full details about order-reduced dynamical Chern-Simons gravity and our methods to simulate black hole spacetimes in this theory are given in~\cite{MashaEvPaper, MashaIDPaper, Okounkova:2017yby}. Here we only briefly  summarize.

The full dCS action takes the form
\begin{align}
\label{eq:dCSAction}
S \equiv \int d^4 x \sqrt{-g} \left( \frac{\mpl^2}{2} R - \frac{1}{2} (\pd \vartheta)^2 - \frac{\mpl}{8} \ell^2 \vartheta \dual RR \right) \,.
\end{align}
The first term is the Einstein-Hilbert action of GR, with the Planck
mass denoted by $\mpl$. The second term in the action is a kinetic
term for the (axionic) scalar field. The third term, meanwhile,
couples $\vartheta$ to spacetime curvature via the parity-odd
Pontryagin density,
\begin{align}
\dual RR \equiv \dual R^{abcd} R_{abcd}\,,
\end{align}
where $\dual R^{abcd} = \frac{1}{2} \epsilon^{abef} R_{ef}{}^{cd}$ is
the dual of the Riemann tensor, and $\epsilon_{abcd} \equiv \sqrt{-g}
[abcd]$ is the fully antisymmetric Levi-Civita tensor~\cite{MTW}. This
interaction is governed by a coupling constant
$\ell$, which has dimensions of length, and physically represents the
length scale below which dCS corrections become relevant.

The equations of motion for $\vartheta$ and $g_{ab}$ have the form
\begin{align}
\square \vartheta \equiv \nabla_a \nabla^a \vartheta = \frac{\mpl}{8} \ell^2 \dual RR\,,
\end{align}
and
\begin{align}
\label{eq:MetricEOM}
    \mpl^2 G_{ab} + \mpl \ell^2 C_{ab} = T_{ab}^\vartheta \,,
\end{align}
where 
\begin{align}
\label{eq:CDefinition}
C_{ab} \equiv \epsilon_{cde(a} \nabla^d R_{b)}{}^c \nabla^e \vartheta + \dual R^c{}_{(ab)}{}^d \nabla_c \nabla_d \vartheta \,,
\end{align}
and $T_{ab}^\vartheta$ is the stress energy tensor for a canonical, massless Klein-Gordon field 
\begin{align}
T_{ab}^\vartheta &= \nabla_a \vartheta \nabla_b \vartheta - \frac{1}{2} g_{ab} \nabla_c \vartheta \nabla^c  \vartheta\,.
\end{align}
Because of $C_{ab}$ in Eq.~\eqref{eq:MetricEOM}, the equation of motion is different from that of a metric in GR sourced by a scalar field. 

$C_{ab}$, as given in Eq.~\eqref{eq:CDefinition}, contains third
derivatives of the metric, and it is thus unknown whether dCS has a
well-posed initial value formulation~\cite{Delsate:2014hba}. But as is typical in the modern treatment of beyond-GR theories
  of gravity, we assume that dCS is a low-energy effective field theory (EFT) of some
  UV-complete theory that is well-posed.  Therefore
we work instead in well-posed order-reduced dCS, in which we perturb the
metric and scalar field about an arbitrary GR spacetime, and obtain
perturbed equations of motion. In particular, we introduce a
  dimensionless formal order-counting parameter $\eps$ which keeps
  track of powers of $\ell^{2}$ (this formal order-counting parameter can later be set to one). 
We then write
\begin{align}
\label{eq:MetricExpansion}
g_{ab} &= g_{ab}^\mathrm{(0)} + \sum_{k = 1}^\infty \eps^k h_{ab}^{(k)} \,, \\
\vartheta &= \sum_{k = 0}^\infty \eps^k \vartheta^{(k)} \,.
\end{align}

Each order in $\eps$ leads to an equation of motion with the same
principal part as GR, and therefore is known to be well-posed at
  each order. Order $\eps^0$ gives the Einstein field equations of
general relativity for $g_{ab}^{(0)}$, the background GR metric,
minimally coupled to a massless scalar $\vartheta^{(0)}$, which
  we can consistently treat as ``frozen out'' and thus set to zero. 
The scalar field is unfrozen at order
$\eps^1$ (cf.~\cite{Okounkova:2017yby}), and it takes the form of a
sourced wave equation
\begin{align}
\label{eq:scalarfield}
    \square^{(0)} \vartheta^{(1)} = \frac{\mpl}{8} \ell^2 \dual RR^{(0)}\,,
\end{align}
where $\square^{(0)}$ is the d'Alembertian operator of the background
and $\dual RR^{(0)}$ is the Pontryagin density of the background. 

Because $\vartheta^{(0)}$ vanishes, there is no correction to the
metric at order $\eps^{1}$. The leading-order dCS correction to the spacetime metric, which will produce the leading-order dCS correction to the gravitational radiation, occurs at order $\eps^2$ (cf.~\cite{Okounkova:2017yby}), and takes the linear form
\begin{align}
\label{eq:SecondOrder}
    \mpl^2 G_{ab}^{(0)} [h_{ab}^{(2)}] = -\mpl \ell^2 C_{ab}^{(1)}  + T_{ab}^{(\vartheta(1))} \,,
\end{align}
where $G_{ab}^{(0)}$ is the linearized
Einstein field equation operator of the background, and 
\begin{align}
\label{eq:KleinGordon}
    T_{ab}^{(\vartheta(1))} &\equiv \nabla_a  {}^{(0)} \vartheta^{(1)} \nabla_b {}^{(0)} \vartheta^{(1)} - \frac{1}{2} g_{ab}^{(0)} \nabla_c {}^{(0)} \vartheta^{(1)} \nabla^c {}^{(0)} \vartheta^{(1)} \,,
\end{align}
where $\nabla_a {}^{(0)}$ denotes the covariant derivative associated with $g_{ab}^{(0)}$. Meanwhile, 
\begin{align}
\label{eq:CTensor}
C_{ab}^{(1)}  &\equiv \epsilon_{cde(a} \nabla^d{}^{(0)} R_{b)}{}^c{}^{(0)} \nabla^e {}^{(0)} \vartheta^{(1)} \\
\nn & \quad + \dual R^c{}_{(ab)}{}^d  {}^{(0)} \nabla_c {}^{(0)} \nabla_d {}^{(0)} \vartheta^{(1)} \,.
\end{align}

To produce beyond-GR gravitational waveforms, our goal is thus to
simultaneously evolve fully nonlinear vacuum Einstein
  equations for $g_{ab}^{(0)}$, Eq.~\eqref{eq:scalarfield} for
  $\vartheta^{(1)}$, and Eq.~\eqref{eq:SecondOrder} for
  $h_{ab}^{(2)}$, to obtain the leading-order dCS correction to the
spacetime metric and corresponding gravitational radiation.

\subsubsection{Scaled variables}
\label{sec:scaling}

Because the evolution equations at some order $\eps^{k}$ are
homogeneous in $\ell^{2k}$,
we can scale out the $\ell$ dependence by defining new variables
\begin{align}
\label{eq:CodeVariables}
h_{ab}^{(2)} \equiv \frac{(\ell/GM)^4}{8} \Delta g_{ab}\,, \; \; \; \vartheta^{(1)} \equiv \frac{\mpl}{8} (\ell/GM)^2 \Delta \vartheta\,
\end{align}
(recall from Sec.~\ref{sec:conventions} that $M$ is the sum of the Christodoulou masses of the background black holes at a given relaxation time~\cite{Boyle:2009vi}). With these substitutions, Eq.~\eqref{eq:scalarfield} becomes
\begin{align}
\label{eq:CodeScalarField}
   \square^{(0)} \Delta \vartheta = \dual RR^{(0)}\,.
\end{align}
Eq.~\eqref{eq:SecondOrder} similarly becomes
\begin{align}
\label{eq:CodeMetric}
G_{ab}^{(0)} [\Delta g_{ab}] = -C_{ab}^{(1)}[\Delta \vartheta] + \frac{1}{8}T_{ab}^{(1)}[\Delta \vartheta]\,.
\end{align}
where $T_{ab}^{(1)}[\Delta \vartheta]$ refers to the Klein-Gordon
stress-energy tensor in Eq.~\eqref{eq:KleinGordon} computed from
$\Delta \vartheta$ instead of $\vartheta^{(1)}$, and
$C_{ab}^{(1)}[\Delta \vartheta]$ similarly refers to the $C$-tensor in
Eq.~\eqref{eq:CTensor} computed with $\Delta \vartheta$ instead of
$\vartheta^{(1)}$.

We thus need to solve Eqs.~\eqref{eq:CodeScalarField} and~\eqref{eq:CodeMetric} only once for each BBH background configuration, and then multiply our results for $\Delta g_{ab}$ and $\Delta \vartheta$ by appropriate powers of $\ell/GM$ and factors of 8 afterward.

\subsection{Evolution} 
 To evolve the first-order dCS metric perturbation, we evolve three systems of equations simultaneously: one for the GR background BBH spacetime, one for the scalar field $\Delta \vartheta$ [cf. Eq.~\eqref{eq:CodeScalarField}] sourced by the background curvature, and one for the metric perturbation $\Delta g_{ab}$ [cf. Eq.~\eqref{eq:CodeMetric}], sourced by the background curvature and $\Delta \vartheta$.  We evolve all variables concurrently, on the same computational domain.

All variables are evolved using the Spectral Einstein Code~\cite{SpECwebsite}, a pseudo-spectral code. The GR BBH background is evolved using a well-posed generalized harmonic formalism, with details given in~\cite{Lindblom2006, Scheel:2008rj, Szilagyi:2009qz, Hemberger:2012jz}. The first-order scalar field is evolved using the formalism detailed in~\cite{Okounkova:2017yby}. Finally, the metric perturbation is evolved using the formalism given in~\cite{MashaEvPaper}, a well-posed perturbed analogue of the generalized harmonic formalism. When evolving the metric perturbation, we have the freedom to choose a perturbed gauge, which we choose to be a harmonic gauge. We give details on perturbed gauge choices in Appendix~\ref{sec:GaugeAppendix}. We use the boundary conditions detailed in~\cite{Cook2004, Rinne:2007ui, Okounkova:2017yby, MashaEvPaper}. 

We use the standard computational domain used for BBH simulations with the Spectral Einstein Code~\cite{SpECwebsite} (such as used in~\cite{Boyle:2019kee}). The computational domain initially has two excision regions (one for each black hole), and the post-merger grid has one excision region (for the final black hole)  (cf. ~\cite{Hemberger:2012jz} for mode details). The outer boundary is chosen to be $\sim 700\,M$. We use adaptive mesh refinement (as detailed in~\cite{Szilagyi:2009qz}), with the background GR variables governing the behavior of the mesh refinement. This is justified, as high gradients in the background will source higher gradients in both the scalar field and the metric perturbation. For all of the evolved variables, in spherical subdomains we filter the top four tensor spherical harmonics, while we use an exponential Chebyshev filter in the radial direction~\cite{Szilagyi:2009qz}. We similarly filter the variables in subdomains with other topologies according the the prescriptions in~\cite{Szilagyi:2009qz}. For the constraint damping parameters (cf.~\cite{Lindblom2006, MashaEvPaper}), we choose the standard values for BBH simulations.

Because the code is pseudo-spectral, we expect roughly exponential convergence with numerical resolution in all of the evolved variables. We specify numerical resolution by choosing adaptive mesh refinement tolerances~\cite{Hemberger:2012jz, Szilagyi:2009qz}; because mesh refinement is based on thresholds, this means that in practice errors decay roughly but not rigorously exponentially---see~\cite{Boyle:2019kee} for further discussion. In~\cite{MashaEvPaper}, we performed detailed tests of the metric perturbation system, showing exponential convergence of evolved variables. We will quote all physical extracted  quantities (cf. Tables~\ref{tab:deltas20} and~\ref{tab:deltas40}) with error bars given by comparing the highest two numerical resolutions.

\subsection{Initial data}

To perform an evolution, we must generate  initial data for the background (metric) fields, the scalar field, and the metric perturbation. The background initial data for a BBH system are given by a constraint-satisfying superposition of black hole metrics in Kerr-Schild coordinates~\cite{Lovelace2009, Ossokine:2015yla}. The scalar field initial data are given by a superposition of slow-rotation solutions~\cite{Okounkova:2017yby, Yunes:2009hc, Yagi:2012ya}. The constraint-satisfying initial data for $\Delta g_{ab}$ are
generated using the methods outlined in~\cite{MashaIDPaper}. For head-on collisions, we start with a separation of $25\,M$, assuming that the contributions to the gravitational radiation and energy flux from times $t \lesssim 25\,M$ are negligible. 

\begin{figure}
  \includegraphics[width=\columnwidth]{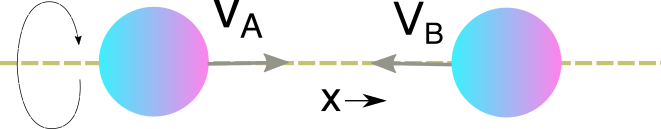}
  \caption{The black hole configurations considered in this study. The two black holes (denoted by spheres) merge along the $x$ axis (as schematically shown by their velocities, $V_A$ and $V_B$). The black holes have equal spins, both oriented in the $+x$ direction, as shown schematically by the gradient on each sphere. The system, as shown by the black arrow on the left, is fully symmetric about the $x$ axis, and additionally has a reflection symmetry $x \to -x$. 
  }
  \label{fig:configurationpp}
\end{figure}

In this study, we will consider axisymmetric configurations where the background spins of the black holes are oriented along $\hat{x}$, the axis along which they are colliding. Moreover, we will choose configurations where the two spins have the same orientation along the axis of collision so that the system has a reflection  symmetry for $x \to -x$ (recall that spin is a pseudo-vector). We illustrate this configuration in Fig.~\ref{fig:configurationpp}. We consider equal mass, equal spin configurations, with dimensionless spins $\chi$ between $0.1$ and $0.8$, in steps of $0.1$. Kerr with $\chi \neq 0$ is not a solution of dCS, and hence the initial configurations will have a non-zero dCS metric perturbation~\cite{Yunes:2009hc}. However, Schwarzschild is a solution of the theory, and hence we do not consider $\chi = 0.0$, as there will be no metric perturbation in that case. 

As a check, we also consider the opposite configuration to Fig.~\ref{fig:configurationpp}, where the spins have opposite orientations. For the equal mass, equal spin systems considered in this study, the final remnant in this case (for all spins) is a Schwarzschild black hole. As Schwarzschild is a solution of dCS, there is zero (to within numerical error) final dCS metric perturbation or scalar field in the spacetime.

\subsection{Wave Extraction}
\label{sec:WaveExtraction}

In the order reduction scheme, $\Psi_4$, the Newman-Penrose scalar measuring the outgoing gravitational radiation is expanded about a GR solution as 
\begin{align}
\label{eq:Psi4Expansion}
    \Psi_4 = \Psi_4^{(0)} + \sum_{k=1}^{\infty} \eps^k \Psi_4^{(k)}\,.
\end{align}
If we substitute the expanded metric given in Eq.~\eqref{eq:MetricExpansion} into the expression for $\Psi_4$ (cf. Appendix \ref{sec:WaveExtractionAppendix}), we can match the terms order-by-order. $\Psi_4^{(1)}$, the first-order correction, will have pieces linear in $h_{ab}^{(1)}$. Recall, however, that $h_{ab}^{(1)} = 0$, so $\Psi_4^{(1)}$ vanishes. $\Psi_4^{(2)}$, the second-order correction, will have pieces quadratic in $h_{ab}^{(1)}$, which will similarly vanish, and pieces linear in $h_{ab}^{(2)}$. Thus, the leading-order correction to the gravitational radiation will be linear in the leading-order correction to the spacetime metric.

In practice, we write [cf.\ Eq.~\eqref{eq:CodeVariables}]
\begin{align}
\Psi_4^{(2)} = \frac{(\ell/GM)^4}{8} \Delta \Psi_4,
\end{align}
and we compute $\Delta \Psi_4$ using the methods detailed in Appendix~\ref{sec:WaveExtractionAppendix}.

Throughout the evolution, we extract $\Psi_4^{(0)}$ and $\Delta \Psi$ on a set of topologically spherical shells using the methods given in~\cite{Taylor:2013zia}. We similarly extract the scalar field $\Delta \vartheta$ radiation on these spherical shells (cf.~\cite{Okounkova:2017yby}). $\Psi_4^{(0)}$ and $\Delta \Psi_4$ are then fit to a power series in $1/r$ (where $r$  is the radius of the spherical shell) and extrapolated to infinity using the methods given in~\cite{Taylor:2013zia, Boyle:2009vi}. We report all of the quantities as $r\Psi_4^{(0)}$ and $r\Psi_4^{(2)}$.

\section{Perturbations to quasi-normal modes}
\label{sec:QNMsec}

Once we have obtained $r\Psi_4^{(0)}$, the background gravitational
radiation, and $r\Psi_4^{(2)}$, the leading order dynamical
Chern-Simons deformation to the gravitational
radiation, we can analyze the quasi-normal mode spectrum.  As discussed in Sec.~\ref{sec:introduction}, head-on
BBH collisions cleanly probe the quasi-normal mode (QNM)
spectrum of the post-merger spacetime. We are thus most interested in
fitting for the QNM spectrum of $r\Psi_4^{(0)}$, and the leading-order deformation to this spectrum in
$r\Psi_4^{(2)}$. A more technical/abstract derivation can be found in
  Appendix~\ref{sec:QNM-formalism}.

\subsection{Quasi-normal modes in general relativity}

A GR QNM waveform takes the form of a superposition of damped sinusoids
\begin{align}
\label{eq:QNMUnexpanded}
r\Psi_4{}(t) = \sum_{l, m, n} \tilde{A}_{(l,m,n)} e^{-i \tilde{\omega}_{(l,m,n)}t}
\, {}_{-2}Y_{(l, m, n)}^{a\omega}\,.
\end{align}
Here, $l$ and $m$ label the spin-weight $-2$ spheroidal harmonic under consideration, while $n$ refers to the overtone, ordered by largest damping time. The quantities $\tilde{A}$ and $\tilde{\omega}$ are the complex amplitude and frequency of the $(l, m,n)$  mode under consideration. For simplicity, we will henceforth omit the $(l,m,n)$ indices on $\tilde{\omega}$ and $\tilde{A}$, and consider each mode separately. We can write $\tilde{\omega}$ in terms of a real frequency, $\omega$, and a damping time, $\tau$, to give
\begin{align}
\label{eq:omega}
    \tilde{\omega} = \omega - i/\tau\,.
\end{align}
Let us similarly write
\begin{align}
    \tilde{A} = A e^{i \theta}\,.
\end{align}
where $A \equiv |\tilde{A}|$ is the norm of $\tilde A$, and $\theta$ is the complex phase of $\tilde A$.
Then we obtain, for a single mode,
\begin{align}
\label{eq:QNMExpanded}
    r\Psi_4 &= A \cos(-\omega t + \theta) e^{-t/\tau} + iA \sin(-\omega t + \theta) e^{-t/\tau}\,.
\end{align}

Since the GR background gravitational radiation is composed of QNMs, we can use the form above to fit for $\Psi_4^{(0)}$ for each mode:
\begin{align}
\label{eq:QNMExpanded}
    r\Psi_4^{(0)} &= A^{(0)} \cos(-\omega^{(0)} t + \theta^{(0)}) e^{-t/\tau^{(0)}} \\
    \nn & \quad + iA^{(0)} \sin(-\omega^{(0)} t + \theta^{(0)}) e^{-t/\tau^{(0)}}\,.
\end{align}
The quantities  $\omega^{(0)}$ and $\tau^{(0)}$ are known from perturbation theory for each $(l, m, n)$~\cite{QNMCode}. Our fit thus determines two free parameters for each mode: $A^{(0)}$, and $\theta^{(0)}$.

\subsection{Perturbed quasi-normal modes}
\label{sec:pert-quasi-norm}
Let us now consider how to fit $r\Psi_4^{(2)}$ after the merger. Note that all fitting is performed in the time domain (cf.~\cite{London:2014cma},~\cite{Giesler:2019uxc}). The QNM frequency, damping time, and amplitude will all be corrected from the background values as
\begin{align}
    \omega &= \omega^{(0)} + \sum_{k = 1}^{\infty} \eps^k \omega^{(k)}\,, \;\;\;\; \tau = \tau^{(0)} + \sum_{k = 1}^{\infty} \eps^k \tau^{(k)}\,, \\
    A &= A^{(0)} + \sum_{k = 1}^{\infty} \eps^k A^{(k)} \,,\;\;\;\;
    \theta = \theta {}^{(0)} + \sum_{k = 1}^{\infty} \eps^k \theta {}^{(k)} \,.
\end{align}
Recall that the leading-order correction to the gravitational radiation is $r\Psi_4^{(2)}$, which is linear in $h^{(2)}_{ab}$ and has a coupling factor of $(\ell/GM)^4$. Thus, the leading-order correction to $\omega$ will be $\omega^{(2)}$, as computed from a linearization of Eq.~\eqref{eq:QNMUnexpanded}, with a coupling factor of $(\ell/GM)^4$. Each mode of $r\Psi_4^{(2)}$, the leading-order dCS correction to the gravitational radiation, will thus be parametrized by and linear in $\{\omega^{(2)}, \tau^{(2)}, A^{(2)}, \theta {}^{(2)}\}$. 

Let us focus on the real part of Eq.~\eqref{eq:QNMExpanded}. Computing the leading-order perturbation to this expression gives us the form
\begin{align}
\mathrm{Re}(r\Psi_4^{(2)}) &= A^{(2)} \cos(-\omega^{(0)} t + \theta^{(0)}) e^{-t/\tau^{(0)}} \\ \nn & \quad - \theta^{(2)} A^{(0)} \sin(-\omega^{(0)} t + \theta^{(0)})  e^{-t/\tau^{(0)}} \\
\nn  & \quad + t \omega^{(2)} A^{(0)} \sin(-\omega^{(0)} t + \theta^{(0)}) e^{-t/\tau^{(0)}} \\
\nn & \quad + t \frac{\tau^{(2)}}{(\tau^{(0)})^2} A^{(0)} \cos(-\omega^{(0)} t + \theta^{(0)}) e^{-t/\tau^{(0)}}\,,
\end{align}
which can be more compactly written as
\begin{align}
s &\equiv \sin(-\omega^{(0)} t + \theta^{(0)}) \,, \\
c &\equiv \cos(-\omega^{(0)} t + \theta^{(0)})  \,, \\
\label{eq:ReDeltaQNM}
\mathrm{Re}(r\Psi_4^{(2)}) &= e^{-t/\tau^{(0)}}  \times \Big[ A^{(2)} c - \theta^{(2)} A^{(0)} s \\
\nn  & \quad + t  A^{(0)} \left(\omega^{(2)} s +  \frac{\tau^{(2)}}{(\tau^{(0)})^2} c\right) \Big]\,.
\end{align}

The imaginary part is similarly modified as

\begin{align}
\label{eq:ImDeltaQNM}
\mathrm{Im}(r\Psi_4^{(2)}) &= e^{-t/\tau^{(0)}}  \times \Big[ A^{(2)} s + \theta^{(2)} A^{(0)} c \\
\nn  & \quad + t  A^{(0)} \left(-\omega^{(2)} c +  \frac{\tau^{(2)}}{(\tau^{(0)})^2} s\right) \Big]\,.
\end{align}

We thus see both an amplitude modification to the background QNM spectrum from the $A^{(2)}$ and $\theta^{(2)}$ terms, and a modification linear in time from the $\omega^{(2)}$ and $\tau^{(2)}$ terms. We fit precisely the functional form in Eqs.~\eqref{eq:ReDeltaQNM} and~\eqref{eq:ImDeltaQNM} to the $r\Psi_4^{(2)}$ obtained from the simulation. Our fit determines four free parameters for each mode: $\{A^{(2)}, \theta^{(2)}, \omega^{(2)}, \tau^{(2)}\}$; the other free parameters $\{A^{(0)}, \theta^{(0)}\}$ in Eqs.~\eqref{eq:ReDeltaQNM} and ~\eqref{eq:ImDeltaQNM} are determined by the fit to $r\Psi_4^{(0)}$ using Eq.~\eqref{eq:QNMExpanded}. \textit{Note that this is different from simply fitting a damped sinusoid to $r\Psi_4^{(2)}$}. 
The presence of a term that behaves as $\sim t e^{-i\tilde{\omega} t}$
beside the ordinary damped sinusoids is a sign of secular breakdown, which is discussed in more detail in Sec.~\ref{sec:secular_regime}.

\subsection{Predictions for particular and homogeneous solutions}
\label{sec:particular}

The metric perturbation $h_{ab}^{(2)}$ satisfies a linear
inhomogeneous differential equation.  Its general solution will be a
linear combination of a homogeneous and particular solution. Shortly after merger, the source driving $h_{ab}^{(2)}$ is constructed from both the
dCS scalar field $\vartheta^{(1)}$ and the nonstationary background
spacetime.  At very late times, when $\vartheta^{(1)}$ settles down to
a stationary configuration, the source for $h_{ab}^{(2)}$ will also be
stationary, sourcing just the stationary deformation
$h_{ab}^{\textrm{Def}}$ away from Kerr, plus any remaining homogeneous
solution.  That homogeneous solution coincides with the GR homogeneous
solution, and thus has the same frequency and decay time as QNMs in GR
(see Appendix~\ref{sec:QNM-formalism} for a more rigorous derivation).

Meanwhile, at earlier times just after merger,
the oscillating $\vartheta^{(1)}$ will
generate a source term for $h_{ab}^{(2)}$
that oscillates at the scalar field's
frequency, and decays at the rate of the scalar field's decay.  Thus
at early times just after merger,
there can be a substantial \emph{particular} solution
with a different frequency and decay time than the late-time
behavior.  We observe this behavior in
Sec.~\ref{sec:observe-part-homog-solut}.

\subsection{Scaling}
\label{sec:tausigns}

Because the simulations (cf. Sec.~\ref{sec:scaling}) are independent of the coupling parameter $\ell/GM$, the resulting waveforms for $\vartheta^{(1)}$ and $r\Psi_4^{(2)}$ have the coupling scaled out. We will thus report our results as
\begin{align}
    &r\vartheta^{(1)} (\ell/GM)^{-2}\,, \;\;\; r\Psi_4^{(2)} (\ell/GM)^{-4} \,, \\
    &\omega^{(2)} (\ell/GM)^{-4} \,, \;\;\; \tau^{(2)} (\ell/GM)^{-4} \,,
\end{align}
and so on. 

Much of the QNM literature reports $\tilde{\omega}$ in terms of its real and imaginary parts, $\tilde{\omega} = \mathrm{Re}(\omega) + i \mathrm{Im}(\omega)$, without invoking a damping time $\tau$. We can transform our results for $\tau$ into $\mathrm{Im}(\omega)$ as
\begin{align}
    \mathrm{Im}(\omega^{(0)}) = -\frac{1}{\tau^{(0)}}\,.
\end{align}
Similarly, given $\tau^{(2)}$, we can perturb the above expression to give
\begin{align}
\label{eq:DeltaTauConversion}
  \mathrm{Im} (\omega^{(2)}) =  \frac{\tau^{(2)}}{(\tau^{(0)})^2}  \,.
\end{align}

\subsection{Mass and spin definitions}

Since $\tau^{(0)}$ and $\omega^{(0)}$ are (by the no-hair theorem) dependent only on $M$ and $\chi$, the mass and dimensionless spin of the final black hole in GR, we should similarly expect $\omega^{(2)}$ and $\tau^{(2)}$ to be dependent on some final mass and final spin. In the full dCS theory, we expect the mass and spin of a dCS black hole to be 
modified with respect to those of a GR black hole (recall that Kerr is not a solution of dCS~\cite{Yunes:2009hc}). The formulae used to compute mass and spin, because they are derived using properties of GR (cf.~\cite{baumgarteShapiroBook}), may themselves be modified in the full dCS theory. If we had access to the full theory, we could parametrize the QNM spectra in terms of $\chi_\mathrm{dCS}$ and $M_\mathrm{dCS}$, as well as $\ell/GM$. Since we are working in an order-reduction scheme, we can instead linearize the formulae used to compute the spin and mass of the final background black hole, and compute the corrected mass and spin. In this study, however, we choose to parametrize the QNM spectra in terms of the Christodoulou mass and dimensionless spin of the final background black hole, which we will call $M_\mathrm{f}$ and $\chi_\mathrm{f}$. 

\subsection{Fitting window}
\label{sec:fitting-window}
When fitting $r\Psi_4^{(0)}$ and $r\Psi_4^{(2)}$, we must be careful about the time window of the post-merger waveform used for the fit. For $r\Psi_4^{(0)}$, if we choose a starting time $t_\mathrm{start}$ too close to merger, then our assumption that each mode can be fit by a function of the form in Eq.~\eqref{eq:QNMUnexpanded} breaks down~\cite{Bhagwat:2017tkm}. The later we choose $t_\mathrm{start}$, the less data (not contaminated by numerical noise) are available to perform the fit. However, in~\cite{Giesler:2019uxc}, the authors found that, when including \textit{overtones}, the post-merger spectrum could be fit with QNMs as early as the peak of the gravitational waveform. We similarly fit enough overtones so that we can faithfully choose $t_\mathrm{start}$ to be the peak of $r\Psi_4^{(0)}$.\footnote{Reference~\cite{Giesler:2019uxc} uses the peak of the gravitational waveform strain as the start of the fit, while we use the peak of $r\Psi_4^{(0)}$.}

\subsection{Practical considerations}
\label{sec:pract-cons}
To perform these linear fits, we use a least-squares method~\cite{scipy}. We fit a sum of overtones to each mode $(l, m)$. We shift $r\Psi_4^{(0)}$ and $r\Psi_4^{(2)}$ to align at the peak of $r\Psi_4^{(0)}$ for each mode. We compute errors in our estimates of the parameters by considering the fitted values for a  medium numerical resolution simulation and a high numerical resolution simulation (for the same initial configuration).

\section{Results}
\label{sec:results}
\subsection{Waveforms}
During each simulation, we extract $r\Psi_4^{(0)}$, the Newman-Penrose scalar measuring the outgoing gravitational radiation of the background spacetime, decomposed into spin-weight $-2$ spherical harmonics labelled by $(l, m)$. Similarly, we extract and decompose $r\Psi_4^{(2)}$, the leading-order dCS correction to the gravitational radiation. Since the computational domain is of finite extent, both quantities are extrapolated to infinity. We additionally extract $\vartheta^{(1)}$, the scalar field, decomposed into spherical harmonics. \textit{In all cases, the spherical harmonics' azimuthal axes are oriented along the collision axis of the black holes, which we will call $\hat{x}$}.\footnote{Note that we decompose into spin-weighted spherical harmonics~\cite{Taylor:2013zia}, not spheroidal harmonics (which do not form a basis), and ignore spherical-spheroidal mode mixing.}

\begin{figure}
  \includegraphics[width=\columnwidth]{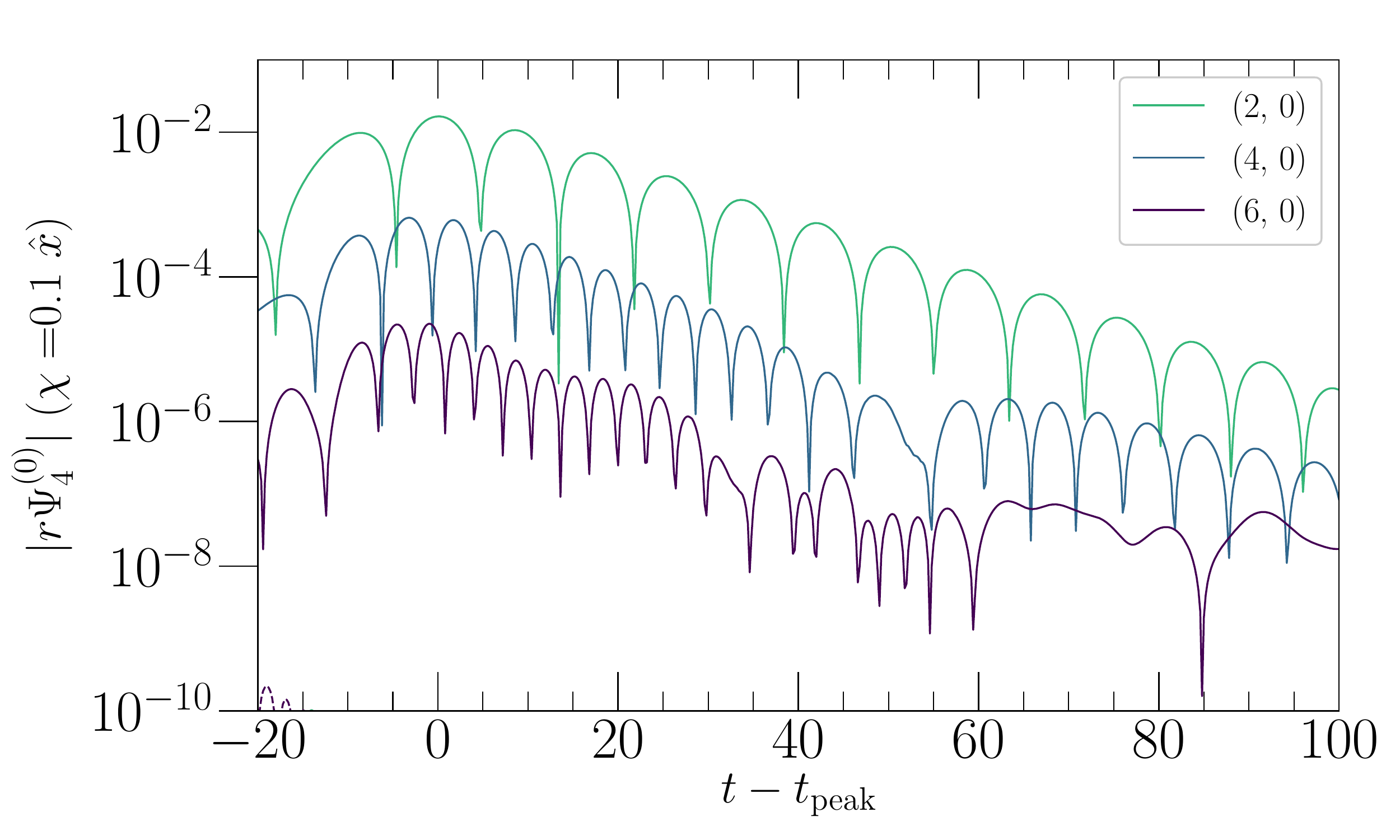}
  \caption{Dominant modes of the background gravitational radiation, shown in terms of the Newman-Penrose scalar $r\Psi_4^{(0)}$ (scaled with radius $r$) for a head-on collision with $\chi = 0.1$ along the axis of the collision (cf. Fig.~\ref{fig:configurationpp}). Each color corresponds to a different mode. For each mode, the solid curves represent the absolute value of the real part of the mode, while there is no power in the imaginary part. We resolve up to the $l = 6$ mode. We choose the reference time $t_\mathrm{peak}$ to correspond to the peak time of the $(2, 0)$ mode of $r\Psi_4^{(0)}$. The data eventually settles to a numerical noise floor (as seen here in the $l = 6$ mode) that exponentially converges towards zero with increasing numerical resolution.}
  \label{fig:Psi4Spin0p1ppX}
\end{figure}

\begin{figure}
  \includegraphics[width=\columnwidth]{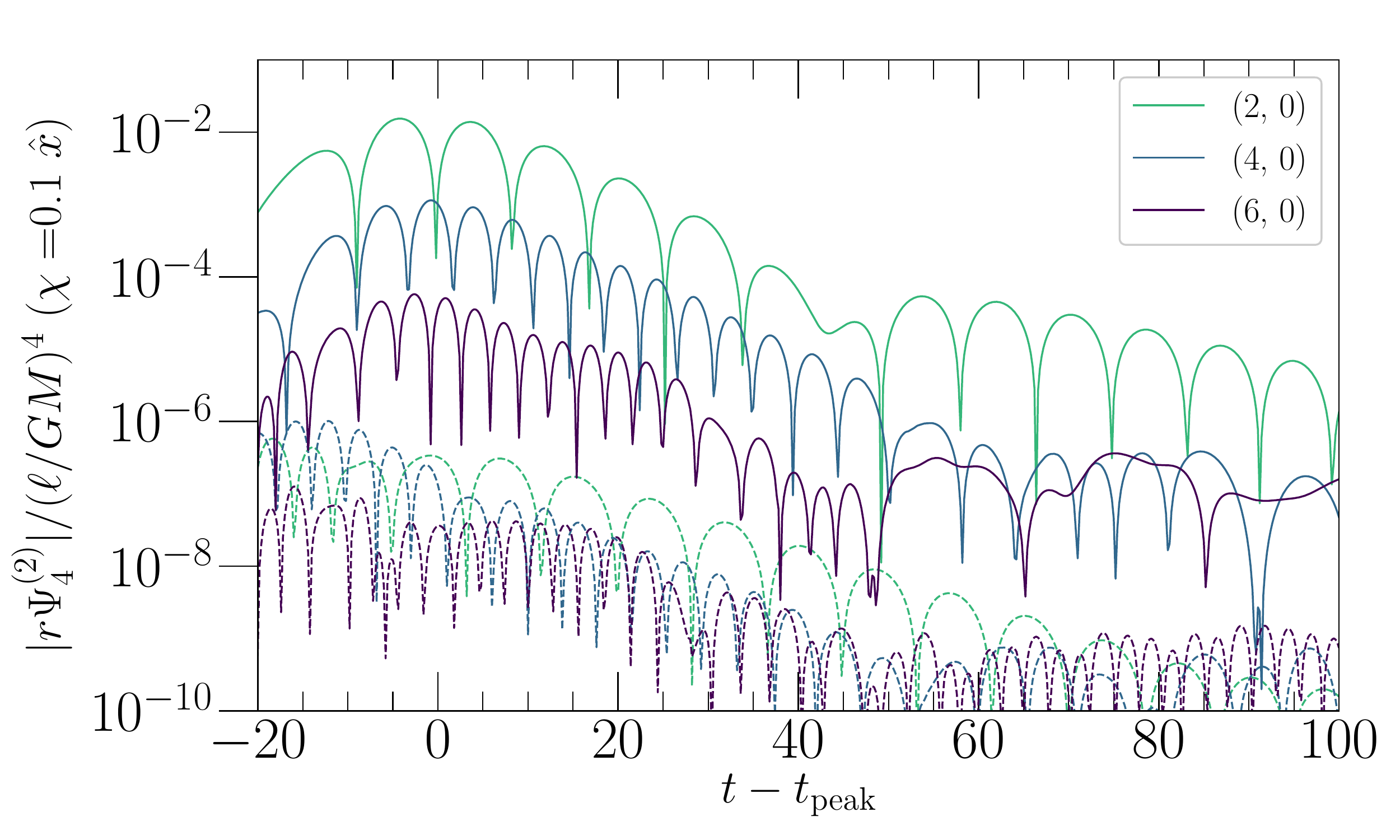}
  \caption{Same as Fig.~\ref{fig:Psi4Spin0p1ppX}, but for the leading-order dCS gravitational radiation, $r\Psi_4^{(2)}$, with the dCS coupling factor $(\ell/GM)^4$ scaled out. Here, there is power in the imaginary part of each mode, which we show with dashed curves. $t_\mathrm{peak}$ is again chosen to correspond to the peak time of the $(2, 0)$ mode of $r\Psi_4^{(0)}$.}
  \label{fig:hPsi4Spin0p1ppX}
\end{figure}

We show the dominant modes of $r\Psi_4^{(0)}$ for a representational case with $\chi =  0.1 \hat{x}$ in Fig.~\ref{fig:Psi4Spin0p1ppX}. We similarly show the dominant modes of $r\Psi_4^{(2)}$ for this configuration in Fig.~\ref{fig:hPsi4Spin0p1ppX}. Recall that the physical gravitational radiation includes a coupling factor $(\ell/GM)^4$, which is scaled out in the numerical computation, and thus we report the waveforms as $(\ell/GM)^{-4} r\Psi_4^{(2)}$. This configuration is axisymmetric about the $\hat{x}$ axis, and thus we expect only $m = 0$ modes to be excited. Since the spins have the same orientation, there is reflection symmetry about the $\hat{y}-\hat{z}$ plane, so we expect only the $l = \mathrm{even}$ modes to be excited. 

\begin{figure}
  \includegraphics[width=\columnwidth]{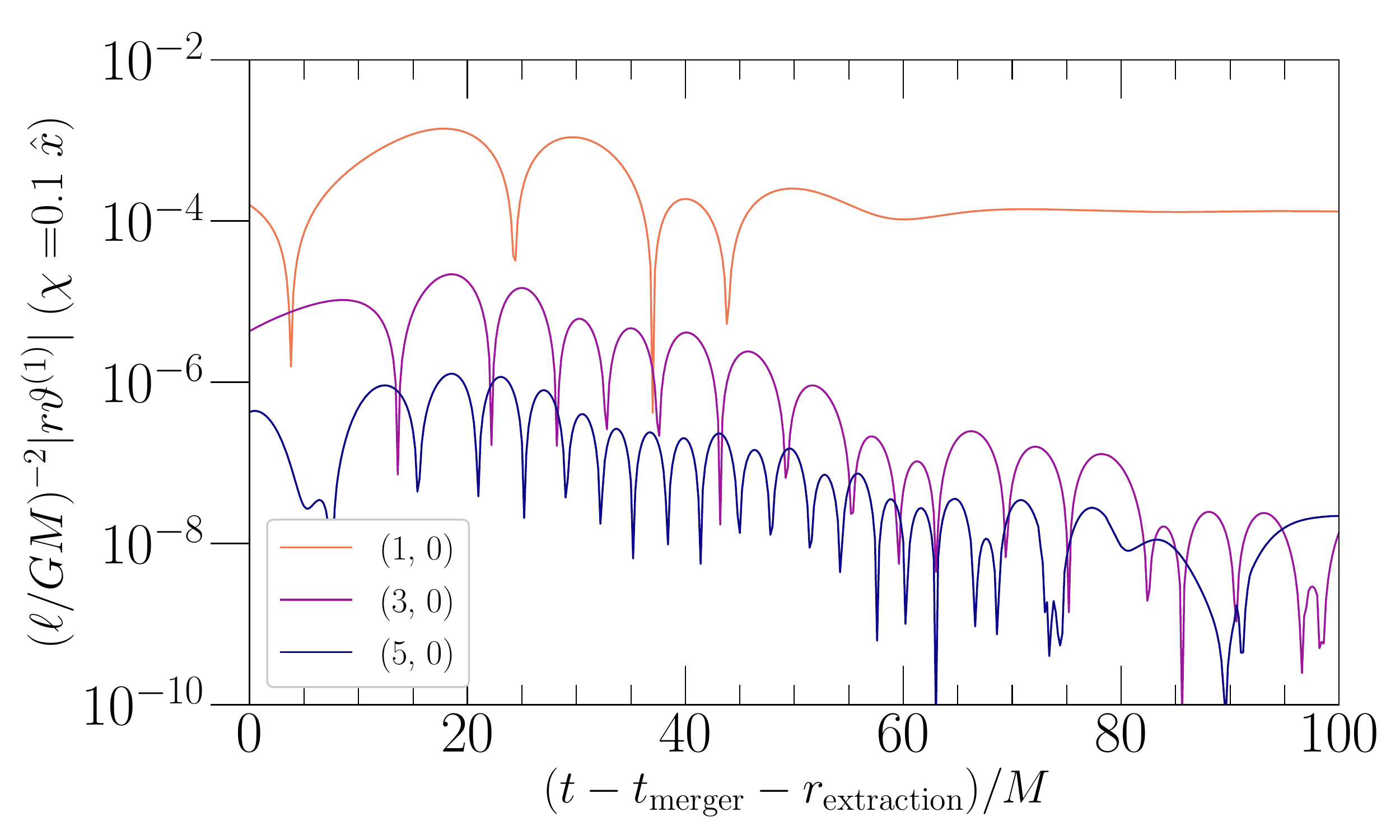}
  \caption{Same as Fig.~\ref{fig:Psi4Spin0p1ppX}, but for the leading-order dCS scalar field $\vartheta^{(1)}$, with the coupling $(\ell/GM)^2$ scaled out. We show the dominantly excited modes of the scalar field, as a function of time relative to merger, corrected by the finite extraction radius $R$ (here shown for $R = 50\,M$). The $(1, 0)$ mode asymptotes to a value corresponding to the dipolar profile of the scalar field around the remnant black hole (we show the results at finite radius to emphasize this point).
  }
  \label{fig:KGPsiSpin0p1ppX}
\end{figure}

Finally, we plot the dominant modes of $\vartheta^{(1)}$, the leading-order dCS scalar field for this configuration, in Fig.~\ref{fig:KGPsiSpin0p1ppX}. Because the scalar field around each black hole takes the form of a dipole oriented around $\hat{x}$ (cf.~\cite{Yagi:2012ya}), and the spins are pointing in the same direction (cf. Fig.~\ref{fig:configurationpp}), we expect power only in the odd $l$ modes. Because of the axisymmetry of the configuration, we expect only the $m = 0$ modes to be excited. We see the $(1, 0)$ mode asymptotes to a value that corresponds to the remnant dipolar profile of the scalar field on the final black hole.

\subsection{Regime of validity}
\label{sec:regime}

\subsubsection{Instantaneous regime of validity}

As discussed in Sec.~\ref{sec:scaling}, the leading-order scalar field $\Delta \vartheta$ and metric perturbation $\Delta g_{ab}$ as computed from the code are independent of the coupling constant $\ell/GM$. In order to make our results physically meaningful, we must multiply the leading-order scalar field by $(\ell/GM)^2$ and the leading-order metric correction by $(\ell/GM)^4$. Similarly, we must multiply the computed leading-order correction to the gravitational radiation, $\Delta \Psi$, by a factor of $(\ell/GM)^4$. 

Recall, however, that the order-reduction scheme is perturbative. The modifications to the spacetime must actually form a convergent perturbation series around GR. We thus require that $g_{ab}$, the background metric, have a larger magnitude than $h_{ab}^{(2)}$ at each point in the spacetime:
\begin{align}
    \| h_{ab}^{(2)} \| \lesssim C \| g_{ab}^{(0)} \|\,,
\end{align}
for some tolerance $C$. This gives an \textit{instantaneous regime of validity}. Following Eq.~\eqref{eq:CodeMetric}, we can compute 
\begin{align}
  \frac{1}{8} (\ell/GM)^4   \| \Delta g_{ab} \| \lesssim C \| g_{ab}^{(0)} \|\,
\end{align}
and hence 
\begin{align}
\label{eq:Validity}
   \left|\frac{\ell}{GM} \right|_\mathrm{max} \lesssim C^{1/4} \left(\frac{8 \left\|g_{ab}\right\|}{\|\Delta g_{ab}\|}\right)_\mathrm{min}^{1/4}
\end{align}
In practice, the ratio is taken point-wise on the computational domain. We choose $C = 0.1$ as a rough tolerance. 

\begin{figure}
  \includegraphics[width=\columnwidth]{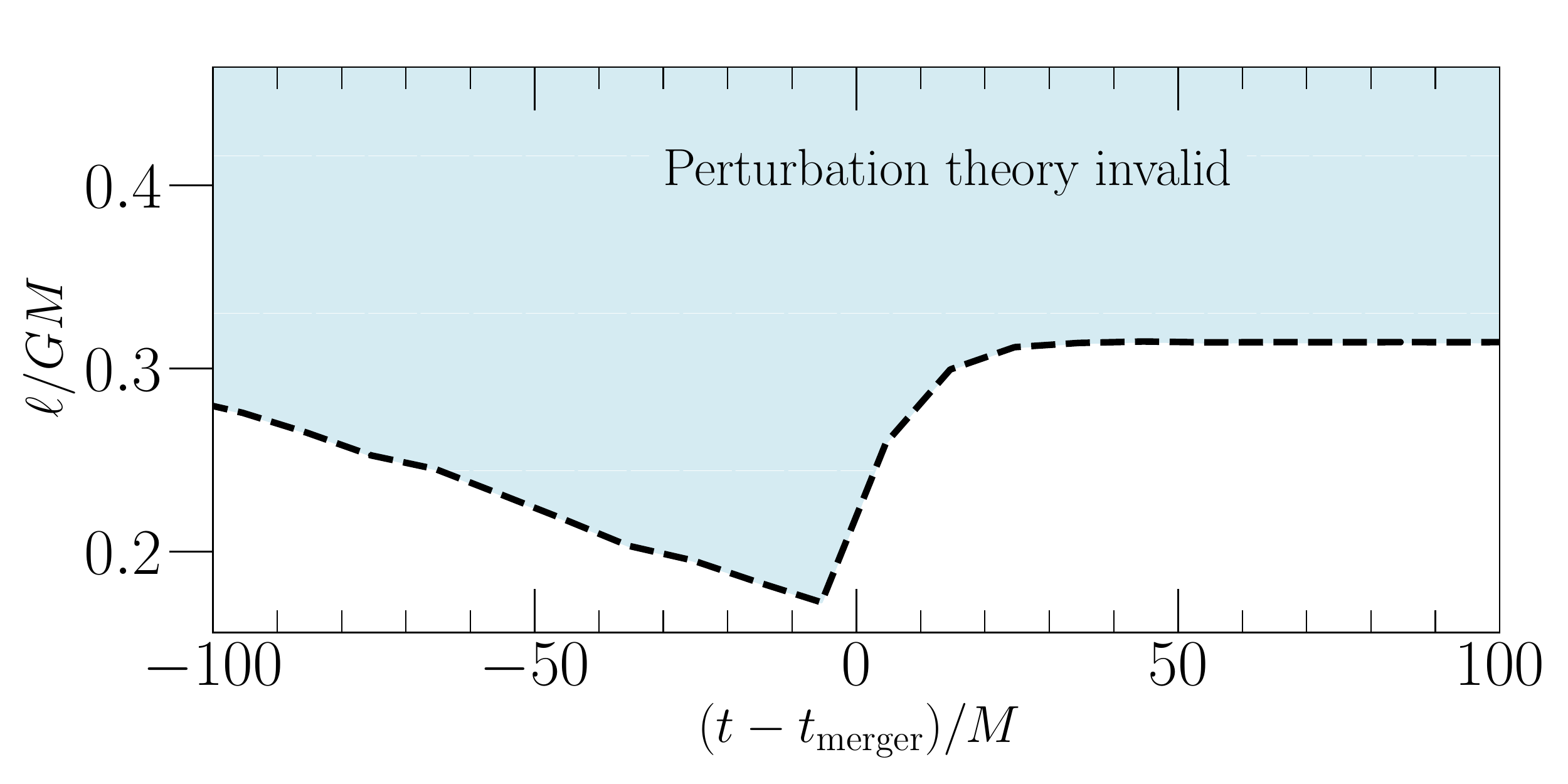}
  \caption{The instantaneous regime of validity for a head-on $\vec\chi = 0.7 \hat{x}$ collision, as a function of coordinate time from merger. On each slice of the simulation, we compute $\ell /GM$, the maximum allowed value of the dCS coupling constant according to Eq.~\eqref{eq:Validity}. The blue region above the dashed curve corresponds to the values of the coupling constant that are not allowed by perturbation theory. Note that this coupling constant appears as $\ell^2$ in the dCS action [cf. Eq.~\eqref{eq:dCSAction}], and as $\ell^4$ in front of the leading-order dCS modification to the gravitational radiation. This figure is in agreement with the estimate in Fig.~5
  of~\cite{Okounkova:2017yby}.
  }
  \label{fig:ValiditySpin0p7ppX}
\end{figure}

We show the regime of validity for a $\vec{\chi} = 0.7 \hat{x}$ head-on collision in Fig.~\ref{fig:ValiditySpin0p7ppX}. $\ell/GM$ takes its smallest allowed value in the strong-field region, outside the apparent horizon of each black hole. We see that closer to merger, where there is power in the metric perturbation, the maximal allowed value of $\ell/GM$ decreases. After merger, the maximal allowed value of $\ell/GM$ increases as the dCS metric perturbation partially radiates away, and the final constant value is governed by the strength of the dCS metric perturbation around the final black hole. 

\begin{figure}
  \includegraphics[width=\columnwidth]{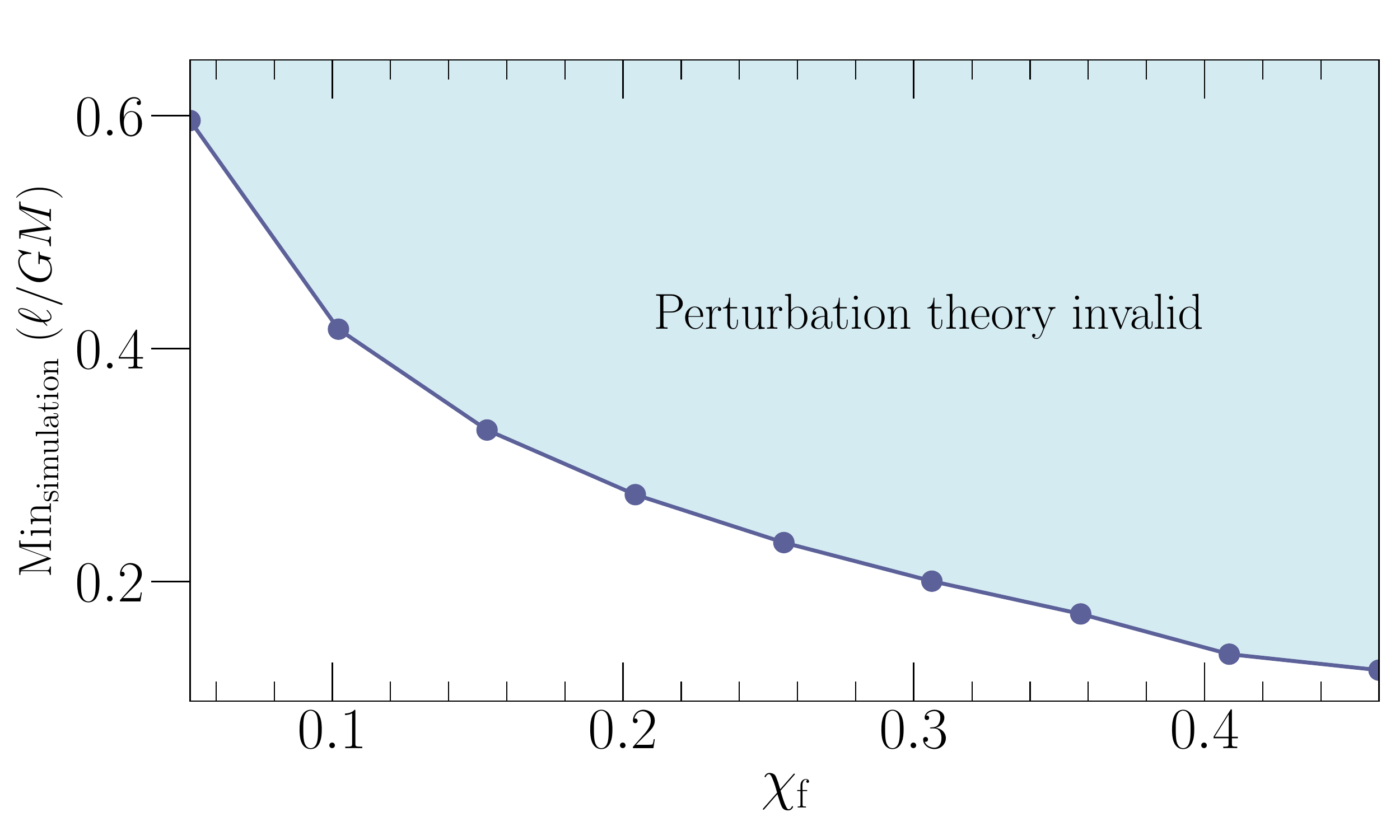}
  \caption{Behavior of the regime of validity with the dimensionless
    spin $\chi_\mathrm{f}$ of the final background black hole.  Each
    black point represents a simulation with a different value of
    $\chi_\mathrm{f}$. We compute the minimum of $\ell /GM$
    [cf.~Eq.~\eqref{eq:Validity}] over time for each simulation. The
    coupling constant achieves its minimum allowed value during the
    merger phase, and thus this regime of validity is a conservative
    estimate. Compare with the earlier results in Fig.~1
    of~\cite{Stein:2014xba} and Fig.~3 of~\cite{MashaIDPaper}.
  }
  \label{fig:MinSpinValidity}
\end{figure}

We show the behavior of the minimum allowed value of $\ell/GM$, over the entire simulation, as a function of final dimensionless spin $\chi_\mathrm{f}$ in Fig.~\ref{fig:MinSpinValidity}. The regime of validity decreases with spin, as the magnitude of $\Delta g_{ab}$ increases with spin. This scaling serves as a proxy for the allowed values of $\ell/GM$ when considering gravitational waveforms. 

\subsubsection{Secular regime of validity}
\label{sec:secular_regime}

In addition to an instantaneous regime of validity, our computations will
also have a \textit{secular} regime of validity. This arises from
our perturbative scheme and is discussed in detail
in~\cite{Okounkova:2017yby}.
In order for the perturbation scheme to remain valid, we want the
perturbation that is going as $\sim t e^{-i \tilde{\omega}^{(0)} t}$ in
Eqs.~\eqref{eq:ReDeltaQNM} and \eqref{eq:ImDeltaQNM} to remain small
relative the background part $\sim e^{-i \tilde{\omega}^{(0)} t}$ in
Eq.~\eqref{eq:QNMExpanded}.  This gives us the condition
\begin{align}
  \left(
    \frac{\ell}{GM}
  \right)^{4}
  \left|
    \tilde{A}^{0} i \tilde{\omega}^{(2)} t e^{-i \tilde{\omega}^{(0)} t}
  \right|
  \lesssim
  \left|
    \tilde{A}^{0} e^{-i \tilde{\omega}^{(0)} t}
  \right|
  \,,
\end{align}
where $|\tilde{\omega}^{(2)}|$ denotes the norm of the leading-order
correction to the complex frequency [cf. Eqs.~\eqref{eq:omega}
and~\eqref{eq:DeltaTauConversion}].
This condition bounds how long the perturbation is valid by the secular time
$t_{\mathrm{sec}}$, roughly
\begin{align}
\label{eq:tsec}
t_\mathrm{sec} =  \min_{lmn} \frac{1}{(\ell/GM)^4|\omega^{(2)}_{lmn}|}\,. 
\end{align}
The (physical) value of $\ell/GM$ thus determines the time window over which the perturbative scheme is appropriate. Since this value is the same for all modes, we must take the most conservative estimate, minimizing the RHS of Eq.~\eqref{eq:tsec} over all modes.

\subsection{Quasi-normal mode fits}
\label{sec:QNM}

We perform the quasi-normal mode fits detailed in Sec.~\ref{sec:QNMsec} to $r\Psi_4^{(0)}$ and $r\Psi_4^{(2)}$. We fit three overtones to each $(l, m)$ mode. For each mode of $r\Psi_4^{(0)}$, we use the perturbation theory results for the corresponding $\omega^{(0)}$, the GR QNM frequency, and $\tau^{(0)}$, the GR damping time~\cite{QNMCode}, and fit for the the QNM amplitudes (cf. Eq.~\eqref{eq:QNMExpanded}). From $r\Psi_4^{(2)}$, we extract $\omega^{(2)}$, the leading-order dCS correction to the QNM frequency, and $\tau^{(2)}$, the leading-order correction to the QNM damping time, as well as the leading-order corrections to the QNM amplitudes (cf. Eqs.~\eqref{eq:ReDeltaQNM} and ~\eqref{eq:ImDeltaQNM}). 

We tabulate all of our fit results in Tables~\ref{tab:deltas20} and~\ref{tab:deltas40}. We quote errors on each of the quantities by comparing the results from the two highest numerical resolutions of the NR simulation.

\begin{figure}
     \centering
     \includegraphics[width=\columnwidth]{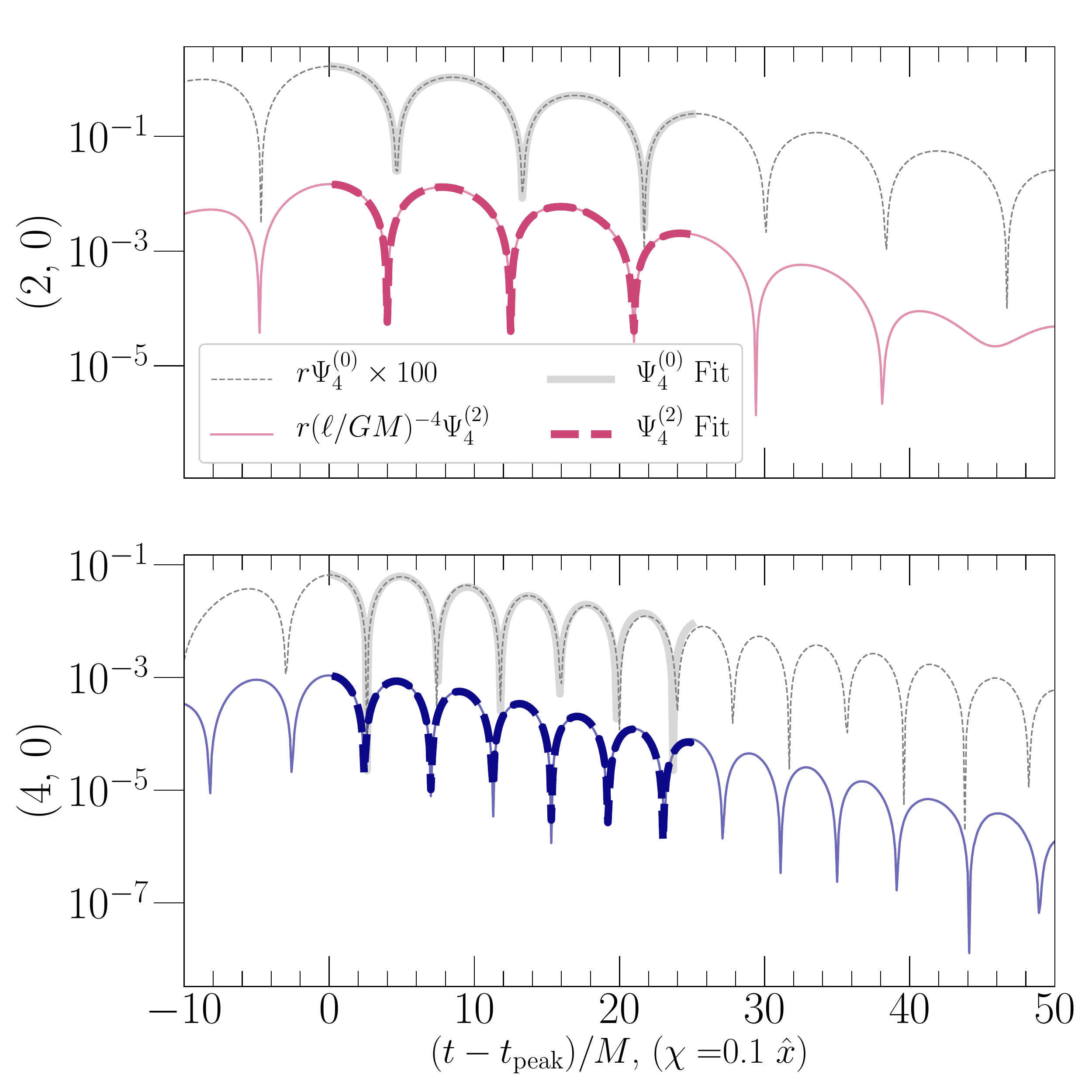}
     \caption{Fits for $r\Psi_4^{(2)}$, the leading-order dCS gravitational radiation, using the formulae in Eqs.~\eqref{eq:ReDeltaQNM} and ~\eqref{eq:ImDeltaQNM}, for a configuration with $\chi = 0.1$ on each hole. Each panel corresponds to one of the dominant modes of the radiation, fit to the three least-damped overtones. The solid colored curves correspond to the real part of $r\Psi_4^{(2)}$. We perform a fit for $r\Psi_4^{(2)}$, shown as
thick dashed colored curves. For reference, we have plotted the real part of $r\Psi_4^{(0)}$ (multiplied by a factor to make it easier to see in this figure) in dashed grey. The QNM fit to $r\Psi_4^{(0)}$ is shown by the solid, thick grey curve.}
     \label{fig:QNMSpin0p1ppX}
\end{figure}

\begin{figure}
     \centering
     \includegraphics[width=\columnwidth]{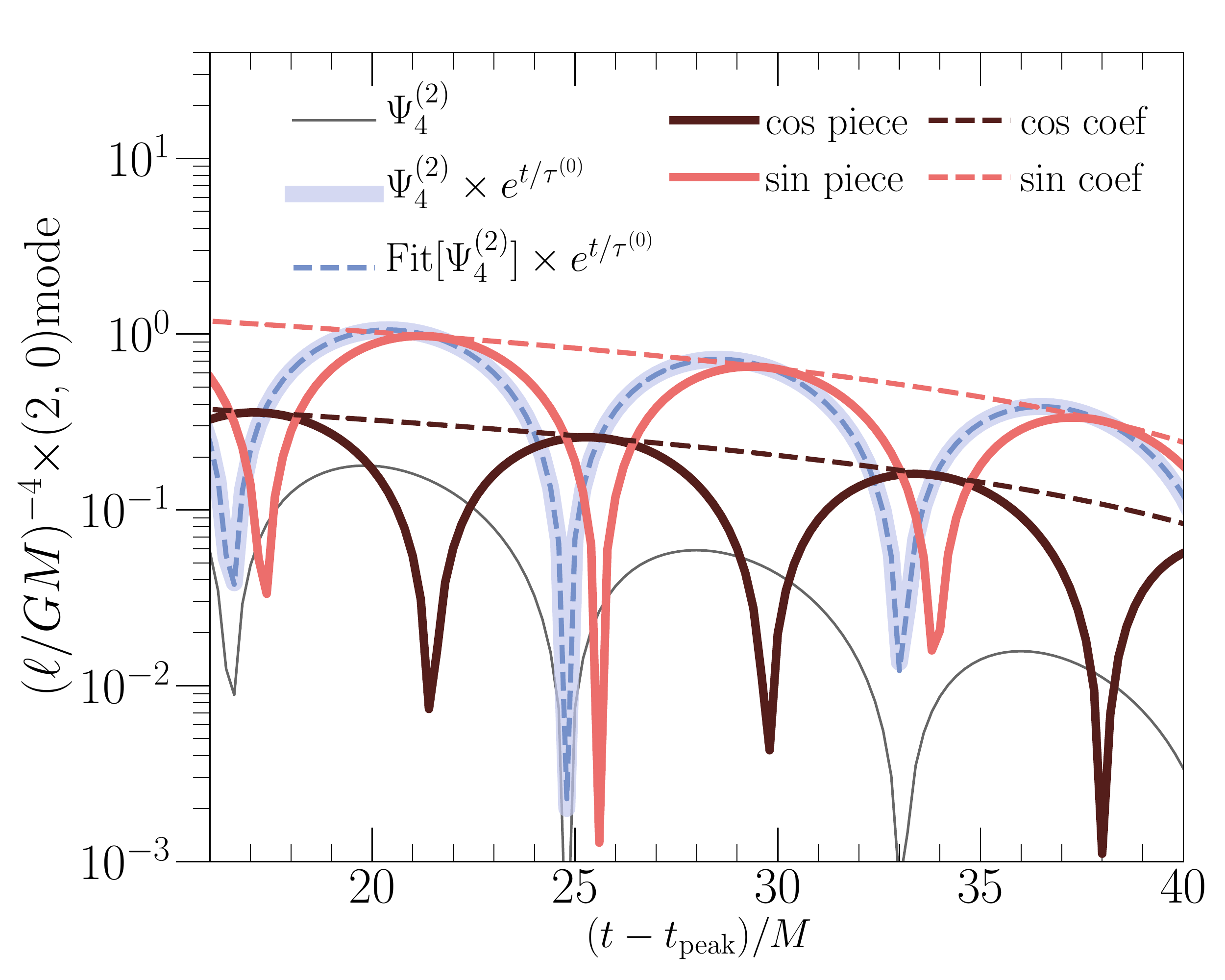}
     \caption{
     The linear-in-time pieces of the $(2,0)$ mode of $r\Psi_4^{(2)}$ (cf. Eqs.~\eqref{eq:ReDeltaQNM} and ~\eqref{eq:ImDeltaQNM}). First, we show $r\Psi_4^{(2)}$ as given by the numerical relativity simulation (solid black). Because $r\Psi_4^{(2)}$ has an overall factor of $e^{-t/\tau^{(0)}}$, we multiply this factor out, showing $r\Psi_4^{(2)} \times e^{t/\tau^{(0)}}$ (thick blue). The resulting waveform then only depends on factors of the form $(a + bt)\cos(\omega^{(0)}t + \theta^{(0)})$, and $(c + dt)\sin(\omega^{(0)}t + \theta^{(0)})$. We individually show these sine and cosine pieces (solid pink and maroon). We then divide $(a + bt)\cos(\omega^{(0)}t+ \theta^{(0)})$ by $\cos(\omega^{(0)}t+ \theta^{(0)})$, leaving only the linear-in-time coefficient $(a + bt)$, and similarly for the sine term (dashed pink and maroon).
     }
     \label{fig:Linearpin0p1ppX}
\end{figure}

We show representative fits in Fig.~\ref{fig:QNMSpin0p1ppX}. We find that for head-on collisions, we can most successfully fit each mode from the peak of the waveform using three overtones. We set $t_\mathrm{end} = 25\,M$ after the peak. We take a closer look at this fit in Fig.~\ref{fig:Linearpin0p1ppX}, where we give an illustration of the linear-in-time behavior of $r\Psi_4^{(2)}$ predicted by Eqs.~\eqref{eq:ReDeltaQNM} and~\eqref{eq:ImDeltaQNM}. We divide the terms of the form $t \cos()$ and $t \sin()$ by $\cos()$ and $\sin()$ respectively, finding that the result is indeed linear with time. 

For each of the configurations, the final Christodolou mass of the background spacetime is $M_\mathrm{f} = 0.9896$. However, $\chi_\mathrm{f}$, the final background spin, varies with configuration, and we include these values in Tables~\ref{tab:deltas20}  and ~\ref{tab:deltas40}.

\begin{figure}
     \centering
     \includegraphics[width=\columnwidth]{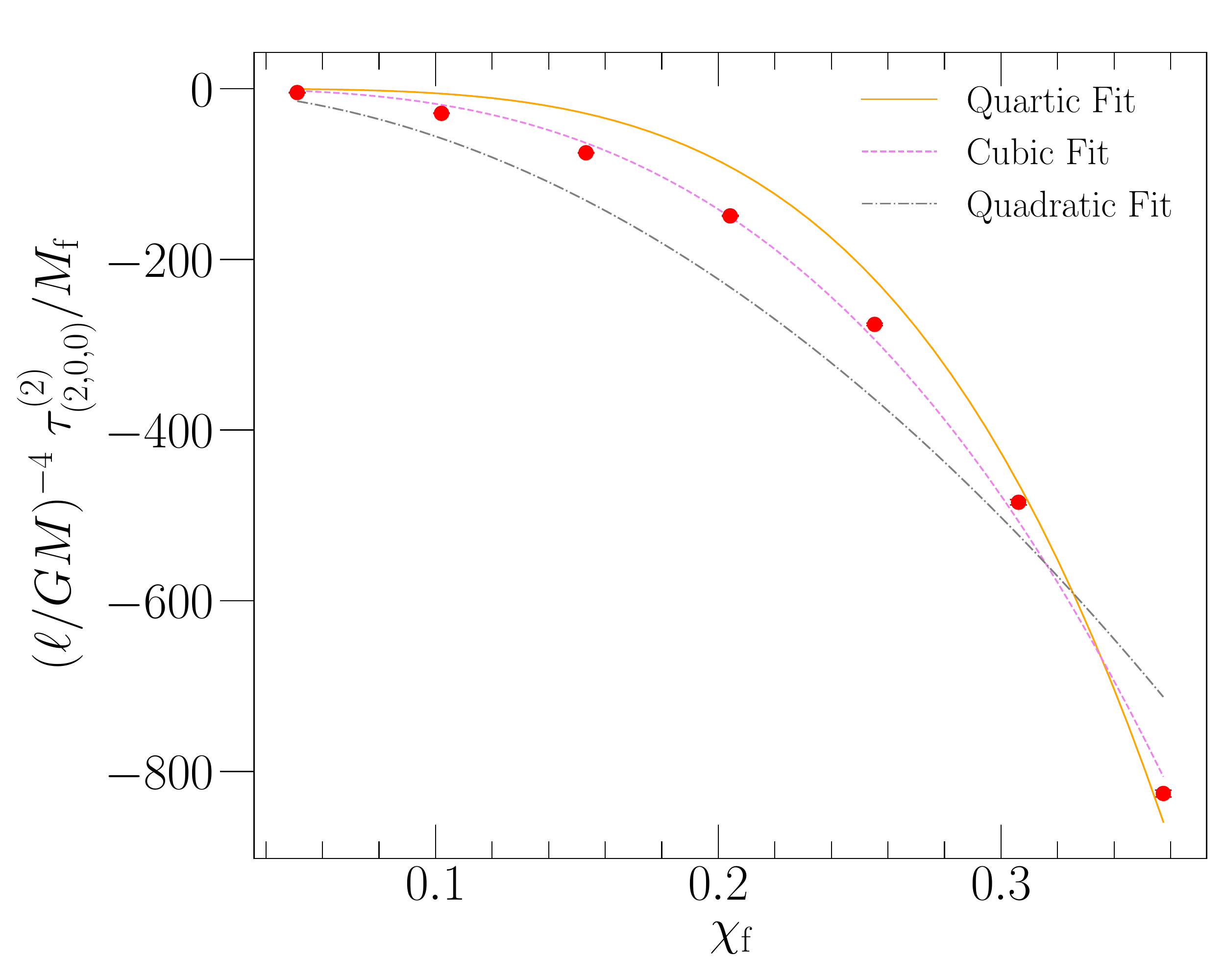}
     \caption{
     Leading-order dCS correction $\tau^{(2)}_{(2, 0, 0)}$ to the QNM damping time $\tau^{(0)}_{(2, 0, 0)}$, plotted as a function of dimensionless spin $\chi_\mathrm{f}$ of the final background black hole. Error bars (barely larger than the plotted points) are computed by considering $\tau^{(2)}$ for numerical simulations with different resolutions (cf.~\ref{sec:QNM}). We see that $\tau^{(2)}$ decreases as a power law with spin. Note that these large values of $\tau^{(2)} (\ell/GM)^{-4}$ must be multiplied by a small, appropriate value of $(\ell/GM)^4$ to have physical meaning.}
     \label{fig:TauSpinFit20}
\end{figure}

\begin{figure}
     \centering
     \includegraphics[width=\columnwidth]{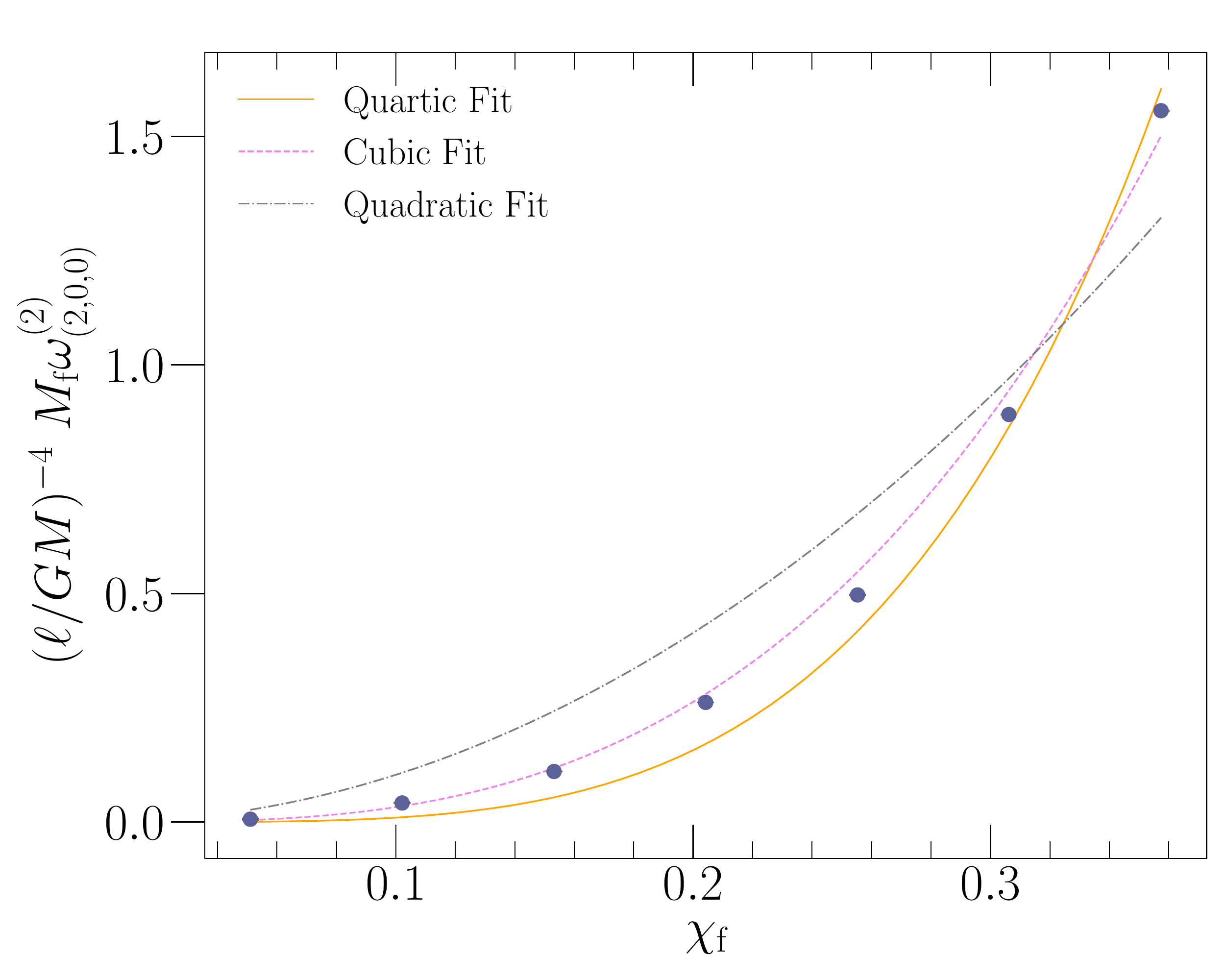}
     \caption{Leading-order dCS correction $\omega^{(2)}_{(2, 0, 0)}$ to the QNM frequency $\omega^{(0)}_{(2, 0, 0)}$, plotted as a function of dimensionless spin $\chi_\mathrm{f}$ of the final background black hole. Error bars (smaller than the plotted points) are computed by considering $\omega^{(2)}$ for numerical simulations with different resolutions (cf.~\ref{sec:QNM}). We see that $\omega^{(2)}$ increases as a power law with spin. Note that these large values of $\omega^{(2)} (\ell/GM)^{-4}$ must be multiplied by a small, appropriate value of $(\ell/GM)^4$ to have physical meaning.
     }
     \label{fig:OmegaSpinFit20}
\end{figure}

\begin{figure}
     \centering
     \includegraphics[width=\columnwidth]{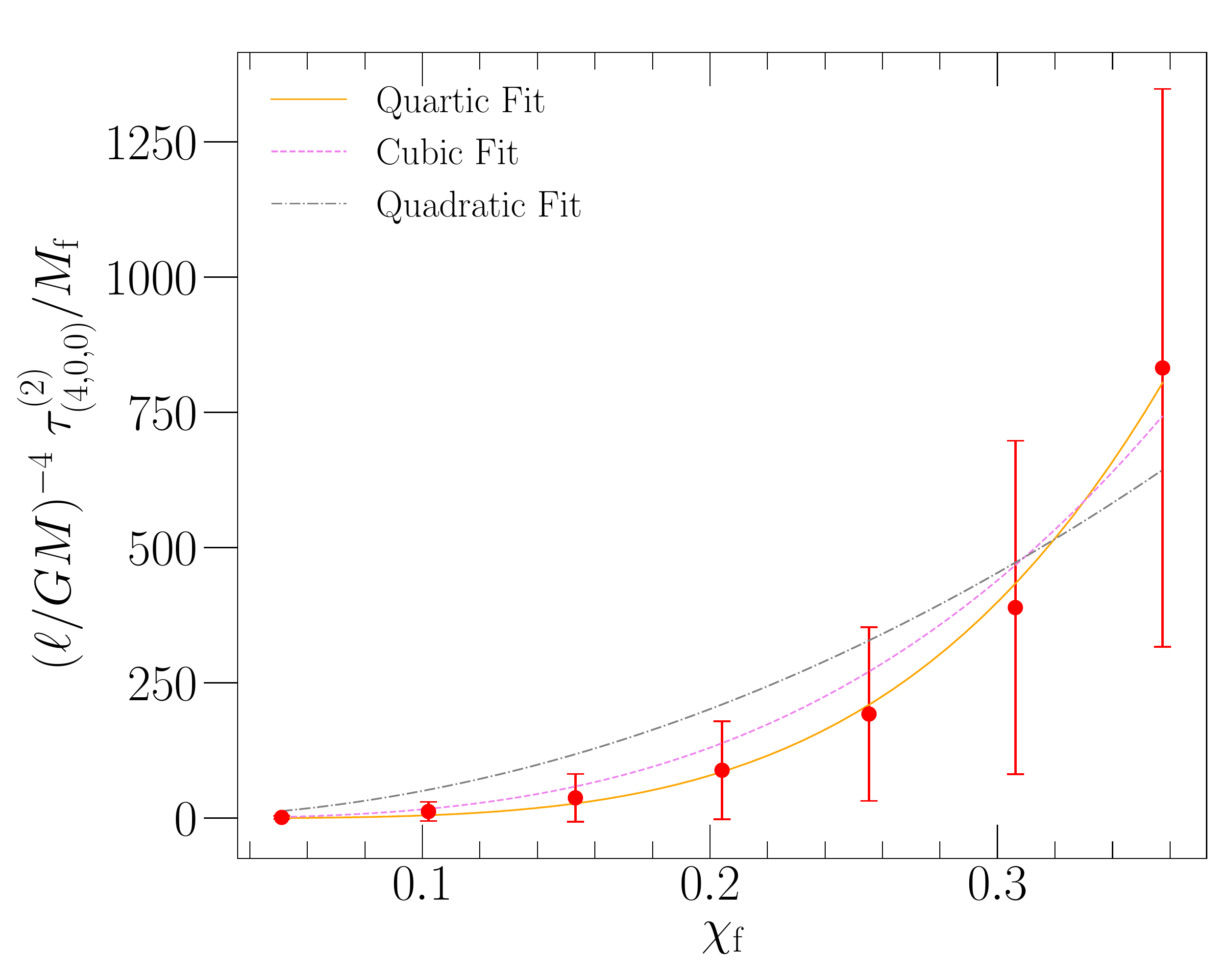}
     \caption{Similar to Fig.~\ref{fig:TauSpinFit20}, but for $\tau^{(2)}_{(4, 0, 0)}$}
     \label{fig:TauSpinFit40}
\end{figure}

\begin{figure}
     \centering
     \includegraphics[width=\columnwidth]{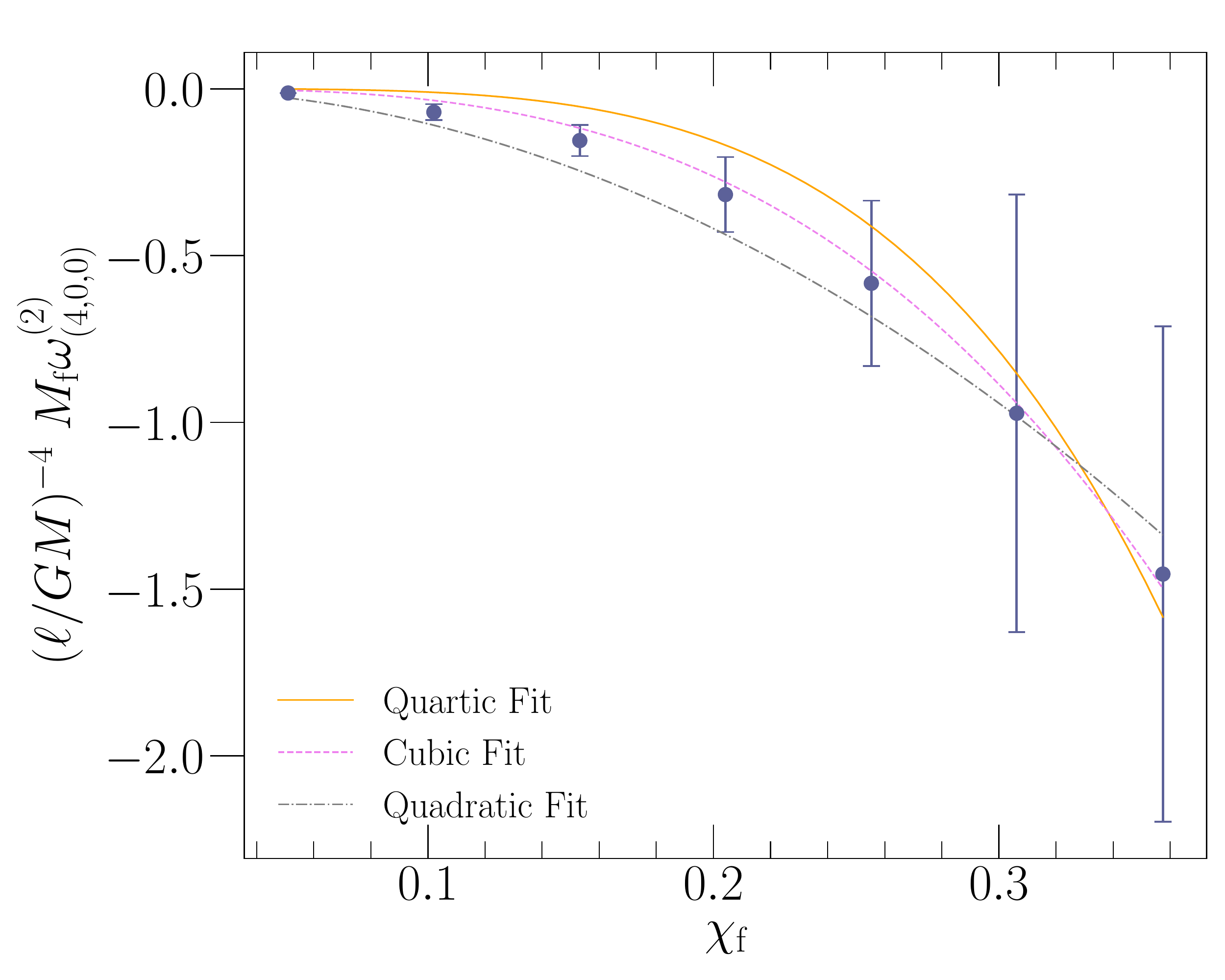}
     \caption{Similar to Fig.~\ref{fig:OmegaSpinFit20}, but for $\omega^{(2)}_{(4, 0, 0)}$}
     \label{fig:OmegaSpinFit40}
\end{figure}

We plot the values of $\tau^{(2)}_{(2,0,0)} (\ell/GM)^{-4}$, the leading-order dCS correction to the damping time of the least-damped $(2,0)$ mode of the gravitational radiation, and $\omega^{(2)}_{(2,0,0)} (\ell/GM)^{-4}$, the leading-order dCS correction to the frequency, as functions of $\chi_\mathrm{f}$ in Figs.~\ref{fig:TauSpinFit20} and~\ref{fig:OmegaSpinFit20}. We see that $\tau^{(2)} (\ell/GM)^{-4}$ and $\omega^{(2)} (\ell/GM)^{-4}$ behave as a power law with spin. This behavior can be expected by considering analytical results in dCS theory. In the slow-rotation approximation, the horizon (and hence the light ring) is modified at quadratic order in spin~\cite{Yagi:2012ya}, while containing no modifications at first order in spin~\cite{Yunes:2009hc}. We additionally plot $ \tau^{(2)} (\ell/GM)^{-4}$ and $ \omega^{(2)}(\ell/GM)^{-4}$ for the $(4,0,0)$ mode in Figs.~\ref{fig:TauSpinFit40} and~\ref{fig:OmegaSpinFit40}. Again, we see that these quantities behave as a power law with spin.

Let us think about the signs of $\tau^{(2)}_{(l,m,n)} (\ell/GM)^{-4}$. In the GR case, given the definition of $\tau^{(0)}$ in Eq.~\eqref{eq:omega}, we always expect $\tau^{(0)}_{(l,m,n)} > 0$ in order for the modes to be exponentially damped (as opposed to exponentially growing). The dCS modifications to the damping time can either be positive or negative. In Fig.~\ref{fig:TauSpinFit20}, for example, we see that for the dominant mode, $(\ell/GM)^{-4}\tau^{(2)}_{(2,0,0)} < 0$, meaning that the damping time is decreased. We can physically think about this as the dCS scalar field removing additional energy from the system, leading to faster damping. We see in Fig.~\ref{fig:TauSpinFit40}, however, that $(\ell/GM)^{-4}\tau^{(2)}_{(4,0,0)} > 0$. 

However, the dCS-modified QNMs should always be exponentially decaying, meaning that we require
\begin{align}
\label{eq:TauSign}
    \tau^{(0)}_{(l, m, n)} + \tau^{(2)}_{(l,m,n)} > 0\,.
\end{align}
Note that the dCS coupling $(\ell/GM)^4$ enters into the above expression. Thus, if we multiply our numerical results for $(\ell/GM)^{-4}\tau^{(2)}_{(l,m,n)}$ (no matter what signs they carry) by the maximum allowed value of $(\ell/GM)^{4}$ to obtain $\tau^{(2)}_{(l,m,n)}$, Eq.~\eqref{eq:TauSign} must hold. We have checked that this is indeed the case (cf. Tables~\ref{tab:deltas20}  and ~\ref{tab:deltas40}).

Let us consider the sources of error in these computations. The resolution of the simulation is the dominant source of error (for example, varying the fitting window as detailed in Sec.~\ref{sec:fitting-window} does not significantly change the results). In each of
Figs.~\ref{fig:TauSpinFit20} and ~\ref{fig:TauSpinFit40}, as well as
the tabulated values in Tables.~\ref{tab:deltas20},
and~\ref{tab:deltas40}, the error bars on a fitted quantity $Q$ are
computed by comparing the value of $Q$ for two simulations with
different numerical resolutions (cf.\ Sec.~\ref{sec:QNM}). The error
bars on the fits for $\tau^{(2)}$, and $\omega^{(2)}$ increase with
$l$, being lowest for the $(2, 0)$ mode, and highest for the $(6, 0)$
mode. Higher modes are more difficult to resolve
numerically~\cite{SpECwebsite, Scheel:2008rj}, and thus it takes
higher resolution for the error bars on the $(6,0)$ mode to decrease
to those on the $(2,0)$ mode at lower resolution. The errors also
increase with the spin of the system. This is because it is more
difficult to resolve higher spin systems
numerically~\cite{Lovelace2008, Lovelace:2010ne}.

\subsection{Particular and homogeneous solutions}
\label{sec:observe-part-homog-solut}
There is interesting behavior later on in the $r\Psi_4^{(2)}$
waveforms. As we can see from e.g. Fig.~\ref{fig:hPsi4Spin0p1ppX},
there is a change in the overall slope that occurs in $r\Psi_4^{(2)}$
around $40\,M$ after the peak time. This change in slope is convergent
with resolution, and is present with and without adaptive mesh
refinement.  Later in the waveform, after this change in slope, both
$r\Psi_4^{(2)}$ and $r\Psi_4^{(0)}$ are well-described by damped
sinusoids, and have the same decay time and frequency. In other words,
at late times $r\Psi_4^{(2)}$ has the same QNM spectrum as
$r\Psi_4^{(0)}$ a GR QNM on a Kerr spacetime (we know from previous work that the resulting GR spacetime of the numerical simulations with this code is Kerr~\cite{Owen:2009sb, Bhagwat:2017tkm}). This suggests that
$r\Psi_4^{(2)}$ switches from being dominantly driven by the dCS
scalar field entering the source term, to being source-free, as
suggested in Sec.~\ref{sec:particular} and discussed further in
Appendix~\ref{sec:part-homog-modes}.  In other words, the early
post-merger dCS waveform correction is dominated by a
\textit{particular solution} of Eq.~\eqref{eq:SecondOrder}, whereas
later it is dominated by a \textit{homogeneous solution}.

\begin{figure}
     \centering
     \includegraphics[width=\columnwidth]{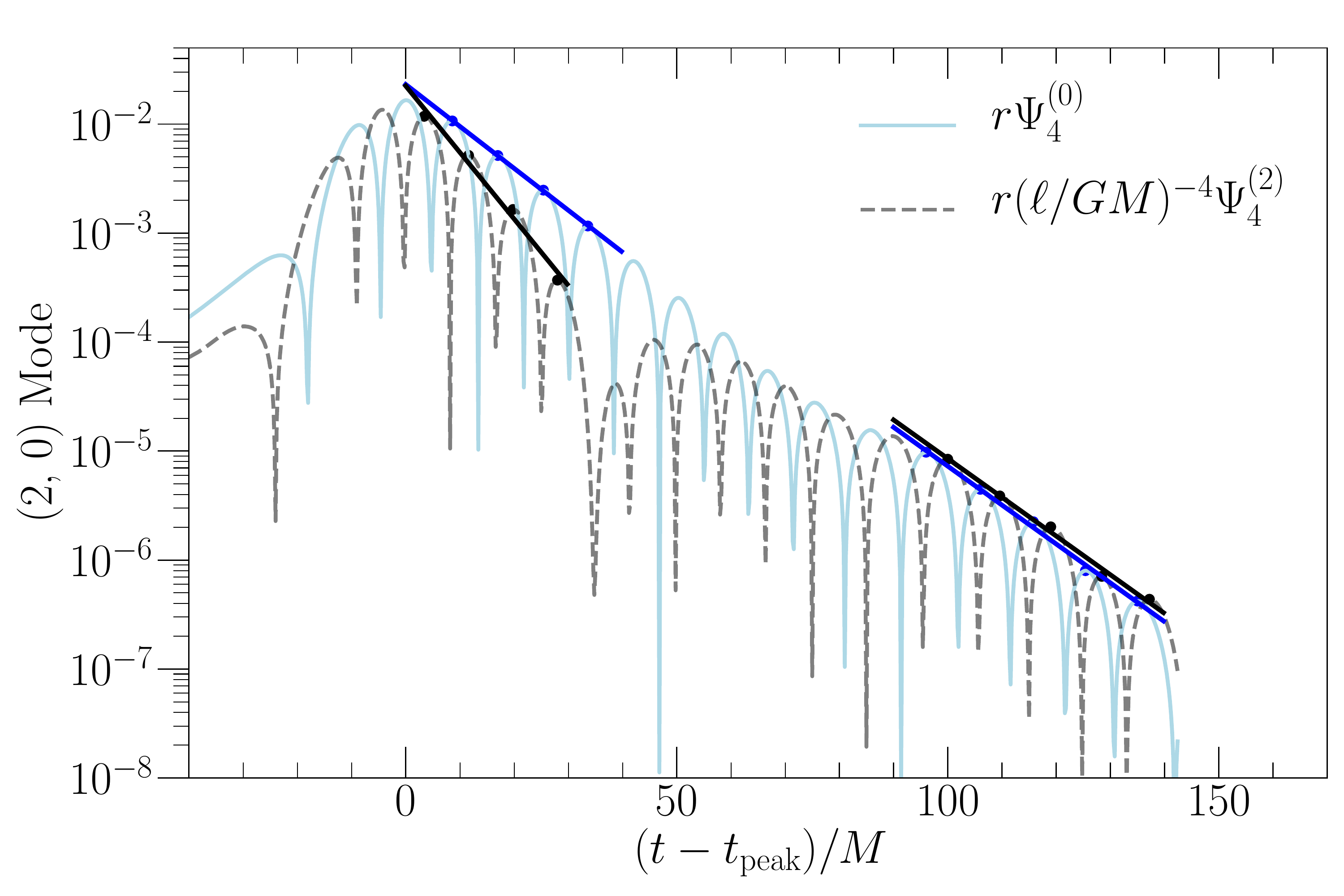}
     \caption{The real parts of $r\Psi_4^{(2)}$ and $r\Psi_4^{(0)}$ for a representative case of $\chi = 0.1\hat{x}$. We fit a line (on this semi-log plot) to the peaks of the gravitational waveform during various stages of the post-merger waveform (black for $r\Psi_4^{(2)}$, blue for $r\Psi_4^{(0)}$). Early on in the waveform, $r\Psi_4^{(2)}$ and $r\Psi_4^{(0)}$ have different damping times, whereas later in the waveform, they have the same damping time. This suggests that at late times, $r\Psi_4^{(2)}$ is well-modeled as a QNM on a pure Kerr background.}
     \label{fig:Slopes}
\end{figure}

We illustrate this behavior schematically in Fig.~\ref{fig:Slopes}.  We
consider the slopes of the logarithms of $r\Psi_4^{(0)}$ and
$r\Psi_4^{(2)}$, which is equivalent to finding a decay time for
each. Note that this is not the same as the perturbed fits for
$r\Psi_4^{(2)}$ given in Eqs.~\eqref{eq:ReDeltaQNM}
and~\eqref{eq:ImDeltaQNM}, which we use to extract $\tau^{(2)}$ and
$\omega^{(2)}$. At late times, the decay times of $r\Psi_4^{(0)}$ and
$r\Psi_4^{(2)}$ are the same. In other words, the late-time leading
order dCS modification to the gravitational radiation is the same as a
GR QNM on Kerr. This behavior is consistent across all spins considered in this study.

We can corroborate this interpretation by looking at the scalar field
in the strong field region, whose dynamics drive the radiative part of
$h_{ab}^{(2)}$. As the scalar field settles down, it is no longer the
dominant source driving $h_{ab}^{(2)}$, and the metric perturbation is
dominantly driven by the Kerr background.  However, making this
interpretation more precise would be tricky: we must keep in mind that
mapping between the strong-field region and a gravitational waveform
at infinity requires utmost care (cf.~\cite{Bhagwat:2017tkm}).

\begin{table*}[htb!]
\begin{center}
  \begin{tabular}{ c | c | c | c | c | c | c | c | c }
    
    \hline
    $\chi_{1,2}$ & $\chi_\mathrm{f}$ & \begin{tabular}{@{}c@{}} $(\ell/GM)^{-4}$ \\ $\omega^{(2)}_{(2,0,0)} M_\mathrm{f}$ 
    \end{tabular}
    &
    \begin{tabular}{@{}c@{}} $(\ell/GM)^{-4}$ \\ $\omega^{(2)}_{(2,0,1)} M_\mathrm{f}$ 
    \end{tabular}
    &
    \begin{tabular}{@{}c@{}} $(\ell/GM)^{-4}$ \\ $\omega^{(2)}_{(2,0,2)} M_\mathrm{f}$ 
    \end{tabular}
    & 
    \begin{tabular}{@{}c@{}} $(\ell/GM)^{-4}$ \\ $\tau^{(2)}_{(2,0,0)} / M_\mathrm{f}$ 
    \end{tabular} & 
    \begin{tabular}{@{}c@{}} $(\ell/GM)^{-4}$ \\ $\tau^{(2)}_{(2,0,1)} / M_\mathrm{f}$ 
    \end{tabular} & 
    \begin{tabular}{@{}c@{}} $(\ell/GM)^{-4}$ \\ $\tau^{(2)}_{(2,0,2)} / M_\mathrm{f}$ 
    \end{tabular} & 
    max
    $(\ell/GM)^4$ \\ \hline

0.1 & 0.05106 & $6.(1) \times 10^{-3}$ & $-1.0(1) \times 10^{-1}$ & $-2.1(3) \times 10^{0}$ & $-4.4(2) \times 10^{0}$ & $8.(1) \times 10^{0}$ & $3.1(4) \times 10^{1}$ &  $1.26\times 10^{-1}$
 \\ \hline
0.2 & 0.1021 & $4.1(2) \times 10^{-2}$ & $-6.9(5) \times 10^{-1}$ & $-1.3(5) \times 10^{1}$ & $-2.8(1) \times 10^{1}$ & $5.4(2) \times 10^{1}$ & $1.9(1) \times 10^{2}$ &  $3.01\times 10^{-2}$
 \\ \hline
0.3 & 0.1532 & $1.1(5) \times 10^{-1}$ & $-1.9(5) \times 10^{0}$ & $-3.5(2) \times 10^{1}$ & $-7.5(7) \times 10^{1}$ & $1.3(1) \times 10^{2}$ & $5.1(1) \times 10^{2}$ &  $1.18\times 10^{-2}$
 \\ \hline
0.4 & 0.2042 & $2.6(5) \times 10^{-1}$ & $-6.1(2) \times 10^{0}$ & $-7.0(2) \times 10^{1}$ & $-1.4(4) \times 10^{2}$ & $3.0(1) \times 10^{2}$ & $1.0(5) \times 10^{3}$ &  $5.68\times 10^{-3}$
 \\ \hline
0.5 & 0.2553 & $4.9(6) \times 10^{-1}$ & $-1.1(6) \times 10^{1}$ & $-1.2(5) \times 10^{2}$ & $-2.7(1) \times 10^{2}$ & $5.6(1) \times 10^{2}$ & $1.9(1) \times 10^{3}$ &  $2.97\times 10^{-3}$
 \\ \hline
0.6 & 0.3062 & $8.9(2) \times 10^{-1}$ & $-2.1(1) \times 10^{1}$ & $-2.2(1) \times 10^{2}$ & $-4.8(3) \times 10^{2}$ & $9.7(2) \times 10^{2}$ & $3.4(2) \times 10^{3}$ &  $1.61\times 10^{-3}$
 \\ \hline
0.7 & 0.3574 & $1.5(2) \times 10^{0}$ & $-4.1(1) \times 10^{1}$ & $-3.8(1) \times 10^{2}$ & $-8.2(4) \times 10^{2}$ & $1.6(2) \times 10^{3}$ & $5.7(3) \times 10^{3}$ &  $8.79\times 10^{-4}$
 \\ \hline

\hline 
  \end{tabular}
\caption{Fitted QNM parameters for each head-on collision configuration considered in this study. All configurations have mass ratio $q = 1$ and final background Christodolou mass $M_\mathrm{f} = 0.9896$. The first column corresponds to the (equal) initial spins of the background black holes, which are oriented in the same direction along the axis of collision (cf. Fig.~\ref{fig:configurationpp}). The second column corresponds to the dimensionless spin $\chi_\mathrm{f}$ of the final background black hole. The third, fourth, and fifth columns correspond to the leading-order dCS correction to the QNM frequencies of the $(2,0,n)$ modes, $\omega^{(2)}_{(2,0)}$ (multiplied by the final background mass, and with the dCS coupling scaled out), for the $n = 0, 1, 2$ overtones. The sixth, seventh, and eighth column similarly correspond to $ \tau^{(2)}_{(2,0,n)}$, the leading-order dCS correction to the QNM damping times (divided by the final background mass and with the dCS coupling scaled out) for the $n = 0, 1, 2$ overtones. We detail our sign convention for $\tau^{(2)}$ in Sec.~\ref{sec:tausigns}. We provide a maximum allowed value of $(\ell/GM)^4$ for each configuration (cf. Sec.~\ref{sec:regime}) in the last column. In order to be physically meaningful, the dCS QNM parameters must be multiplied by this factor. We have  checked that adding $\mathrm{max} ((\ell/GM)^4) \tau^{(2)}_{(2,0,n)}$ to the GR background solution $\tau^{(0)}_{(2,0,n)}$ does not change the sign (meaning that all modes are decaying).}
\label{tab:deltas20}
\end{center}
\end{table*}


\begin{table*}[htb!]
\begin{center}
  \begin{tabular}{ c | c | c | c | c | c | c | c | c }
    
    \hline
    $\chi_{1,2}$ & $\chi_\mathrm{f}$ & \begin{tabular}{@{}c@{}} $(\ell/GM)^{-4}$ \\ $\omega^{(2)}_{(4,0,0)} M_\mathrm{f}$ 
    \end{tabular}
    &
    \begin{tabular}{@{}c@{}} $(\ell/GM)^{-4}$ \\ $\omega^{(2)}_{(4,0,1)} M_\mathrm{f}$ 
    \end{tabular}
    &
    \begin{tabular}{@{}c@{}} $(\ell/GM)^{-4}$ \\ $\omega^{(2)}_{(4,0,2)} M_\mathrm{f}$ 
    \end{tabular}
    & 
    \begin{tabular}{@{}c@{}} $(\ell/GM)^{-4}$ \\ $\tau^{(2)}_{(4,0,0)} / M_\mathrm{f}$ 
    \end{tabular} & 
    \begin{tabular}{@{}c@{}} $(\ell/GM)^{-4}$ \\ $\tau^{(2)}_{(4,0,1)} / M_\mathrm{f}$ 
    \end{tabular} & 
    \begin{tabular}{@{}c@{}} $(\ell/GM)^{-4}$ \\ $\tau^{(2)}_{(4,0,2)} / M_\mathrm{f}$ 
    \end{tabular} & 
    max
    $(\ell/GM)^4$ \\ \hline

0.1 & 0.05106 & $-1.2(5) \times 10^{-2}$ & $4.(2) \times 10^{-2}$ & $8.5(1) \times 10^{-2}$ & $1.(3) \times 10^{0}$ & $-7.(1) \times 10^{-1}$ & $-6.(3) \times 10^{-1}$ &  $1.26\times 10^{-1}$
 \\ \hline
0.2 & 0.1021 & $-6.(2) \times 10^{-2}$ & $2.(2) \times 10^{-1}$ & $4.8(8) \times 10^{-1}$ & $1.(1) \times 10^{1}$ & $-5.(1) \times 10^{0}$ & $-4.2(7) \times 10^{0}$ &  $3.01\times 10^{-2}$
 \\ \hline
0.3 & 0.1532 & $-1.5(4) \times 10^{-1}$ & $5.(5) \times 10^{-1}$ & $1.2(2) \times 10^{0}$ & $3.(4) \times 10^{1}$ & $-1.2(4) \times 10^{1}$ & $-1.0(3) \times 10^{1}$ &  $1.18\times 10^{-2}$
 \\ \hline
0.4 & 0.2042 & $-3.(1) \times 10^{-1}$ & $1.(1) \times 10^{0}$ & $2.5(6) \times 10^{0}$ & $8.(9) \times 10^{1}$ & $-2.8(1) \times 10^{1}$ & $-2.2(6) \times 10^{1}$ &  $5.68\times 10^{-3}$
 \\ \hline
0.5 & 0.2553 & $-5.(2) \times 10^{-1}$ & $1.(2) \times 10^{0}$ & $4.(1) \times 10^{0}$ & $1.(1) \times 10^{2}$ & $-5.6(2) \times 10^{1}$ & $-4.(1) \times 10^{1}$ &  $2.97\times 10^{-3}$
 \\ \hline
0.6 & 0.3062 & $-9.(6) \times 10^{-1}$ & $2.(3) \times 10^{0}$ & $7.(2) \times 10^{0}$ & $3.(3) \times 10^{2}$ & $-1.0(1) \times 10^{2}$ & $-8.(1) \times 10^{1}$ &  $1.61\times 10^{-3}$
 \\ \hline
0.7 & 0.3574 & $-1.4(7) \times 10^{0}$ & $2.(8) \times 10^{0}$ & $1.2(5) \times 10^{1}$ & $8.(5) \times 10^{2}$ & $-1.8(1) \times 10^{2}$ & $-1.6(3) \times 10^{2}$ &  $8.79\times 10^{-4}$
 \\ \hline
  \end{tabular}
\caption{Same as Table~\ref{tab:deltas20}, but for the $(4,0,n)$ modes.}
\label{tab:deltas40}
\end{center}
\end{table*}


\section{Implications for testing general relativity}
\label{sec:testingGr}

Let us now discuss this work in the context of testing GR with gravitational wave observations. Suppose that we were to observe a post-merger gravitational wave, given by $r\Psi_4$. We can check the consistency of this waveform with GR, and with dCS.

Let us first suppose that $\ell = 0$, meaning that there is no
modification from GR. In GR, assuming the no-hair theorem is true,
the frequency and damping time for each mode of ringdown should be
parametrized purely by the mass, $M_\mathrm{f}$, and spin,
$\chi_\mathrm{f}$, of the final black hole. Given two observed modes, we have access to four quantities --- two frequencies and two damping times. We can check that the fitted $\omega_{(l,m,n)}$ and $\tau_{(l,m,n)}$ are consistent with the predicted GR values for $M_\mathrm{f}$ and $\chi_\mathrm{f}$~\cite{Berti:2015itd, Gossan:2011ha, TheLIGOScientific:2016src, Kamaretsos:2011um, Yunes:2016jcc}. 

If the values are not consistent with GR, meaning that there is a shift away from the predicted GR frequencies and damping times, we can check whether the fitted $\omega_{(l,m,n)}$ and $\tau_{(l,m,n)}$ agree with the leading-order dCS-corrected frequencies and damping times computed using the methods in this study.

\subsection{Checking non-degeneracy: projected}

Let us consider whether it is possible in principle to measure dCS
corrections $\omega_{(l,m,n)}^{(2)}$ and $\tau_{(l,m,n)}^{(2)}$ from a
given waveform signal, or instead whether these quantities are
degenerate with the GR values $\omega_{(l,m,n)}^{(0)}$ and
$\tau_{(l,m,n)}^{(0)}$ corresponding to a different remnant mass and
spin. Consider the 4-dimensional parameter space $\mathbb P$ of
$\{M\omega^{(0)}, \tau^{(0)}/M, M\omega^{(2)}, \tau^{(2)}/M\}$ for a
given mode. GR solutions exist purely in the 2-dimensional submanifold $\mathbb S_{\mathrm{GR}}$ specified by $\{\omega^{(2)}=0, \tau^{(2)}=0\}$, with coordinates $\{M_\mathrm{f}, \chi_\mathrm{f}\}$ on the manifold. Suppose at some point $(M_1, \chi_1)$ on $\mathbb S_{\mathrm{GR}}$, we introduce a dCS deviation with some coupling $(\ell/GM)^4$. In other words, we will have
\begin{align}
\label{eq:OmegaDeviation}
\omega(\chi_1, M_1) &= \omega^{(0)}(\chi_1, M_1) + (\ell/GM)^4 \Delta \omega (\chi_1, M_1)\,, \\
\label{eq:TauDeviation}
\tau(\chi_1, M_1) &= \tau^{(0)}(\chi_1, M_1) + (\ell/GM)^4 \Delta \tau (\chi_1, M_1)\,,
\end{align}
where we have explicitly written out the dependence on the coupling
constant with $\Delta \omega \equiv (\ell/GM)^{-4} \omega^{(2)}$ and
$\Delta \tau \equiv (\ell/GM)^{-4} \tau^{(2)}$, which can be read
  off of Figs.~\ref{fig:TauSpinFit20}--\ref{fig:OmegaSpinFit40}.
If dCS modifications and GR are degenerate, then this modification will move purely within $\mathbb S_{\mathrm{GR}}$. However, if dCS modifications and GR are non-degenerate, then the new point will be off $\mathbb S_{\mathrm{GR}}$ in $\mathbb P$ and the dCS modifications will form a 3-dimensional submanifold of $\mathbb P$, $\mathbb S_{\mathrm{dCS}}$,  with coordinates $\{M_\mathrm{f}, \chi_\mathrm{f}, \ell/GM\}$.

\begin{figure}
     \centering
     \includegraphics[width=\columnwidth]{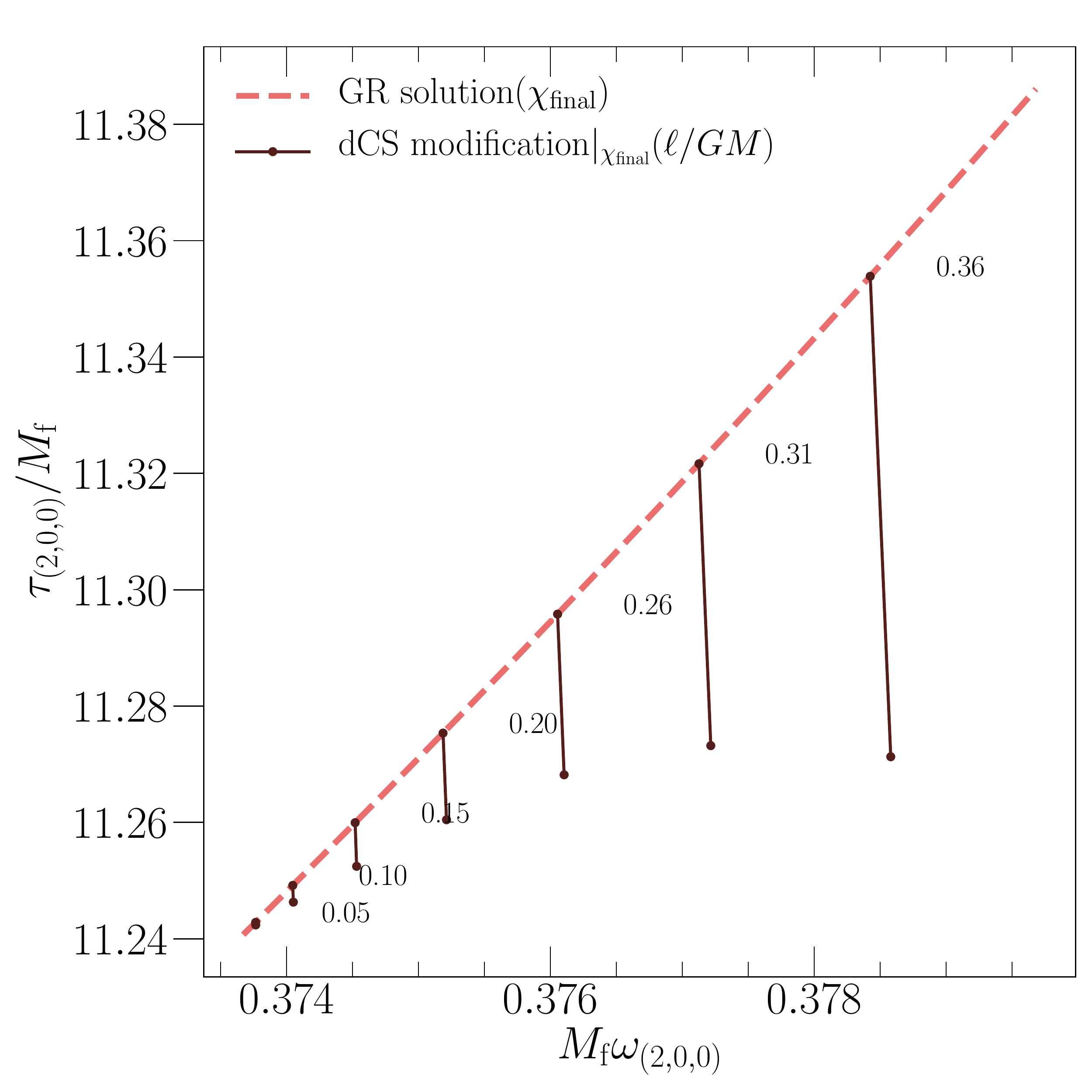}
     \caption{Probing degeneracy of GR and dCS-corrected QNM spectra. We show the values of $M_\mathrm{f} \omega$ and $\tau/M_\mathrm{f}$ for the $(2,0,0)$ mode of the post-merger gravitational radiation. If there is no dCS modification, i.e. $\ell = 0$, then for fixed final mass $M_\mathrm{f}$, the GR QNM solutions form a curve (dashed pink) parametrized by $\chi_\mathrm{f}$ in the plane. For each $\chi_\mathrm{f}$ we introduce a dCS modification using the $\omega^{(2)}$ and $\tau^{(2)}$ that we have computed in this study. This modification depends on the coupling parameter $(\ell/GM)$, and thus forms a line (recall that the dependence on $(\ell/GM)$ is purely \textit{linear}) parametrized by $(\ell/GM)$  in the $M_\mathrm{f}\omega  -  \tau/M_\mathrm{f}$ plane (solid maroon). Each such line on the plot is labeled by the value of the final dimensionless spin. Here we choose a conservative maximum value of $(\ell/GM)^4 = 10^{-4}$ for each spin. We see that this modification does not purely lie along the GR solution, and hence GR and dCS-corrected QNM spectra are non-degenerate. 
     }
     \label{fig:OmegaTauDegeneracy20}
\end{figure}

\begin{figure}
     \centering
     \includegraphics[width=\columnwidth]{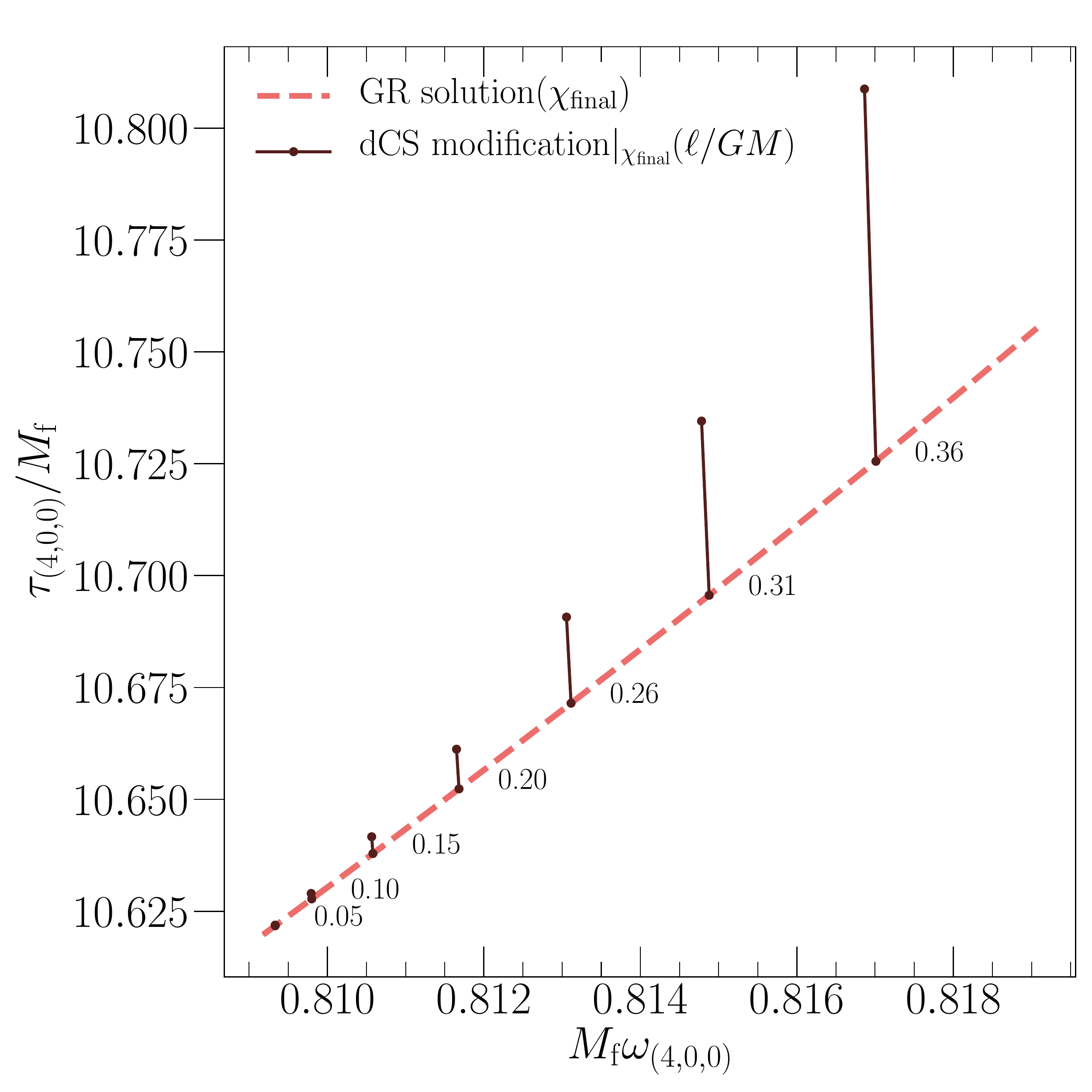}
     \caption{Same as Fig.~\ref{fig:OmegaTauDegeneracy20}, but for the $(4,0,0)$ mode of the gravitational radiation. 
     }
     \label{fig:OmegaTauDegeneracy40}
\end{figure}

Let us now consider this statement in the context of our numerical results. For simplicity, let us first consider holding $M_\mathrm{f}$ constant in the comparisons. In Fig.~\ref{fig:OmegaTauDegeneracy20}, we plot values of $M_\mathrm{f}\omega_{(2,0,0)}$ and $\tau_{(2,0,0)}/M_\mathrm{f}$ for various spins. We similarly plot $M_\mathrm{f}\omega_{(4,0,0)}$ and $\tau_{(4,0,0)}/M_\mathrm{f}$ in Fig.~\ref{fig:OmegaTauDegeneracy40}. For $\ell = 0$, we can use perturbation theory to compute the values of $\omega = \omega^{(0)}$ and $\tau = \tau^{(0)}$ in GR. Holding $M_\mathrm{f}$ fixed, the GR solutions form a curve $L_\mathrm{GR}$ in the $\omega-\tau$ plane, parametrized by $\chi_\mathrm{f}$. 

Now let us introduce $\ell \neq 0$. For each simulation that we have performed, with a given $\chi_\mathrm{f}$ (recall all of the $M_\mathrm{f}$ are equal), we compute $\omega(\chi_\mathrm{f})$ and $\tau(\chi_\mathrm{f})$ via Eqs.~\eqref{eq:OmegaDeviation} and~\eqref{eq:TauDeviation} using our results for 
$\Delta \omega = (\ell/GM)^{-4} \omega^{(2)}$ and $\Delta \tau = (\ell/GM)^{-4} \tau^{(2)}$. This computation requires specifying a value of $(\ell/GM)$. If we vary $(\ell/GM)$ over an allowed range (cf. Sec.~\ref{sec:regime}), for each $\chi_\mathrm{f}$ we obtain a curve $L_\mathrm{dCS}(\chi_\mathrm{f})$ in the $M_\mathrm{f}\omega - \tau/M_\mathrm{f}$ plane parametrized by $(\ell/GM)$.

If the dCS corrections to the quasi-normal mode spectrum were degenerate with GR, then $L_\mathrm{dCS}(\chi_\mathrm{f})$ would lie purely along $L_\mathrm{GR}$. In other words, the resulting QNM spectrum for $\chi_\mathrm{f}$ would be degenerate with that of GR for some other spin $\chi'$. However, we see in Figs.~\ref{fig:OmegaTauDegeneracy20} and~\ref{fig:OmegaTauDegeneracy40} that in all cases $L_\mathrm{dCS}(\chi_\mathrm{f})$ does \textit{not} lie purely along $L_\mathrm{GR}$, meaning that the QNM spectra are non-degenerate. This in turn means that dCS modifications to QNM spectra can in principle be observed (in the limit of infinite signal-to-noise ratio). Note that in this analysis, we have held $M_\mathrm{f}$ fixed, since all of the head-on collision simulations in this paper have the same final black hole mass. 


\subsection{Checking non-degeneracy: full case}

We can perform a more rigorous analysis, checking for full degeneracy, rather than the simpler check that holds $M_\mathrm{f}$ fixed. Let us think about the 3-dimensional space $\mathbb{F}$ with coordinates $\{\chi, M, \eps^2\}$ (where $\eps$ is our dCS order-reduction parameter). Suppose we observe $k$ QNMs, which gives us $2k$ quantities ($\omega$ and $\tau$ for each mode). Let $\mathbb{Q}$ be the $2k$-dimensional space with these coordinates.

Let us consider the map $\phi: \mathbb{F} \to \mathbb{Q}$, which maps each set of parameters $\{\chi, M, \eps^2\}$ to the QNM values. The image $\phi(\mathbb{F})$ will form a 3-dimensional submanifold of $\mathbb{Q}$, and the tangent space of the image will be spanned by the pushforwards of $\{\pd/\pd \chi, \pd/\pd M, \pd/\pd \eps^2\}$. That is, $\{\phi_* \pd/\pd \chi, \phi_* \pd/\pd M, \phi_* \pd/\pd \eps^2\}$. 

Non-degeneracy in this context means that the dimension of the span of
$\{\phi_* \pd/\pd \chi, \phi_* \pd/\pd M, \phi_* \pd/\pd \eps^2\}$ is
3.  This can be checked by looking at the rank of the $2k \times 3$
dimensional (Jacobian) matrix
\begin{align}
  J \equiv  \begin{bmatrix} 
    \begin{pmatrix} \\
      \phi_* \frac{\pd}{\pd \chi}\\
      ~
    \end{pmatrix}&
    \begin{pmatrix} \\
      \phi_* \frac{\pd}{\pd M}\\
      ~
    \end{pmatrix}&
    \begin{pmatrix} \\
      \phi_* \frac{\pd}{\pd \eps^{2}}\\
      ~
    \end{pmatrix}
\end{bmatrix}  \,.
\end{align}

Let us consider how to evaluate this matrix, working near $\eps^2 = 0$ for each $\chi_\mathrm{f}$ and $M_\mathrm{f}$ for which we have performed a head-on collision. Suppose we are considering some mode with QNM frequency $\omega_{lmn}$ and damping time $\tau_{lmn}$. This will give us two rows in the matrix $J$.

Let us first compute 
\begin{align}
    &\frac{\pd}{\pd \chi} \omega_{lmn} = \frac{1}{M_\mathrm{f}} \frac{\pd}{\pd \chi} (\omega_{lmn} M_\mathrm{f}) \,, \\
    &\frac{\pd}{\pd \chi} \tau_{lmn} = M_\mathrm{f} \frac{\pd}{\pd \chi} (\tau_{lmn} / M_\mathrm{f})
\end{align}
This can be done by computing the values of $\omega_{lmn}
M_\mathrm{f}$ and $\tau_{lmn}/M_\mathrm{f}$ from perturbation
theory~\cite{QNMCode}, varying only $\chi$ around
$\chi_\mathrm{f}$, and then taking a numerical derivative. We work
with a step-size of $10^{-10}$, which is the default precision
of~\cite{QNMCode}.

Now let us compute the $\pd/\pd M$ column. For fixed $\chi_\mathrm{f}$ and $\ell^4 = 0$, the dependence on $M$ is 
\begin{align}
    \frac{\pd}{\pd M} \omega_{lmn} &= \frac{1}{M} \frac{\pd}{\pd M} (\omega_{lmn} M) = \frac{\omega_{lmn}}{M} \,, \\
    \frac{\pd}{\pd M} \tau_{lmn} &= M \frac{\pd}{\pd M} \left(\frac{\tau_{lmn}}{M}\right) = -\frac{\tau_{lmn}}{M}\,.
\end{align}
We evaluate these expressions at $M = M_\mathrm{f}$.

Finally, for the last column, for fixed $\chi_\mathrm{f}, M_\mathrm{f}$, we have
\begin{align}
    \frac{\pd}{\pd \eps^2} \omega_{lmn} &= \frac{1}{M_{f}} (\ell/GM)^{-4}M_{f}\omega_{lmn}^{(2)} \,, \\
    \frac{\pd}{\pd \eps^2} \tau_{lmn} &= M_{f} (\ell/GM)^{-4}\frac{\tau_{lmn}^{(2)}}{M_{f}}\,,
\end{align}
where $(\ell/GM)^{-4}M_{f}\omega_{lmn}^{(2)}$ and
$(\ell/GM)^{-4}\tau_{lmn}^{(2)}/M_{f}$ are the quantities we compute
from our numerical fits.

We put the matrix $J$ together with these values. Note that the
$\omega$ rows all have a factor of $1 / M_\mathrm{f}$, while the
$\tau$ rows have a factor of $M_\mathrm{f}$. Since each row
  is homogeneous in a power of $M_\mathrm{f}$, we can divide through without changing the rank of the matrix. 
We evaluate the rank of this matrix using a singular-value
decomposition (SVD)~\cite{scipy}. For all values of
$\{\chi_\mathrm{f}, M_\mathrm{f}\}$ in our head-on collisions dataset,
we find that the rank of $J$ is 3. The lowest singular value is
$10^{-2}-10^{-1}$, while the condition numbers (the 2-norm, computed
from the SVD) are of order $10^{3}$. Therefore, we conclude that dCS corrections to ringdowns have
  a distinct observational signature, as they are not degenerate
  with changing GR parameters of final mass and spin.


\section{Conclusion}
\label{sec:conclusion}
In this study, we have produced the first beyond-GR BBH gravitational waveforms in full numerical relativity for a higher-curvature theory. We have considered head-on collisions of BBHs in dynamical Chern-Simons gravity. While these are not likely to be astrophysically relevant configurations, they serve as a proof of principle of our ability to produce beyond-GR waveforms~\cite{MashaEvPaper}. Future work in this program thus involves adding initial orbital angular momentum to the system and producing beyond-GR gravitational waveforms for inspiraling systems. We have previously evolved the leading order dCS scalar field for an inspiraling BBH background~\cite{Okounkova:2017yby}, and can use our (fully-general) methods given in~\cite{MashaIDPaper} and~\cite{MashaEvPaper} to produce initial data for and evolve an inspiraling BBH system. 

We have  also studied modifications to the post-merger BBH head-on collision QNM spectra. We found that at leading order, the damping time of each QNM receives a modification that increases with the spin of the final black hole as a power law. The frequency of each QNM receives a similar modification. These modifications are not degenerate with GR. 

When performing inspiraling BBH simulations, we can repeat the analysis outlined in this paper to learn about the dCS modification to the QNM spectrum of an astrophysically relevant system. These results can then be applied to beyond-GR tests of BBH ringdowns~\cite{TheLIGOScientific:2016src, Yunes:2016jcc}. In particular, the investigation of modification to overtones is useful for an analysis of the form detailed in~\cite{Isi:2019aib}. Note that the GR-dCS non-degeneracy results found in this paper assume infinite signal to noise ratio. Thus, future work also includes checking degeneracy in the presence of gravitational wave detector noise. Inspiraling simulations will also allow us to perform even more powerful tests of GR using \textit{full} inspiral-merger-ringdown waveforms, thus taking advantage of the entire gravitational wave signal.


\section*{Acknowledgements}

This work was supported in part by the Sherman Fairchild Foundation, and NSF grants PHY-1708212 and PHY-1708213 at Caltech and PHY-1606654 at Cornell. Computations were performed using the Spectral Einstein Code~\cite{SpECwebsite}. All computations were performed on the Wheeler cluster at Caltech, which is supported by the Sherman Fairchild Foundation and by Caltech.

\appendix

\section{Formalism for QNMs beyond GR}
\label{sec:QNM-formalism}

In this Appendix we give an abstract formalism for QNM modeling in
theories beyond GR.  For simplicity we will present perturbative
expansions with leading power $\varepsilon^{1}$, but the
generalization to the behavior of dCS (where the leading metric
correction is at $\mathcal{O}(\varepsilon^{2})$) is straightforward.

\subsection{QNMs in GR}
\label{sec:QNMs-GR}

Under the final state conjecture~\cite{Penrose:1969pc,
  Chrusciel:2012jk, klainerman}, the result of a merger of two Kerr
black holes will uniquely be a perturbed Kerr BH, with the
perturbation decaying with time.  Therefore, the waveform
after merger is typically modeled using linear black hole perturbation
theory.  That is, we treat the post-merger metric as
\begin{align}
  g_{ab} = g_{ab}^{K} + \xi h^{Q}_{ab} + \mathcal{O}(\xi^{2}) \,,
\end{align}
where $g_{ab}^{K}$ is the Kerr metric, and $\xi$ is a small
formal order-counting parameter.  The full metric satisfies the
nonlinear Einstein equations up to order $\xi^{2}$, yielding the
linear partial differential equation (PDE) that $h_{ab}$ satisfies,
\begin{align}
  \label{eq:met-pert}
  G_{ab}[g_{cd}^{K} + \xi h^{Q}_{cd} ] = \mathcal{O}(\xi^{2})
  \,,
  \quad\Longrightarrow\quad
  G^{(1)}_{ab}[h^{Q}_{cd}] = 0 \,,
\end{align}
where $G^{(1)}[\cdot]$ is the Einstein operator linearized about the
Kerr background.
Notice that QNMs are homogeneous solutions to this linear PDE.

In practice, the metric perturbation equations~\eqref{eq:met-pert} of
GR are intractable, whereas curvature perturbations for the Weyl
scalar $\Psi_{0}$ and $\Psi_{4}$ can be decoupled, yielding the
Teukolsky equation~\cite{Teukolsky:1972my, Teukolsky:1973ha},
which we will denote as
\begin{align}
  \mathcal{T}\,\Psi_{4} = 0 \,.
\end{align}
In the Teukolsky formalism, QNMs are still homogeneous solutions.
As Teukolsky showed, this partial differential equation is amenable to
separation of variables.  The most general homogeneous solution, at
large $r$, is a linear combination
\begin{align}
  \label{eq:rPsi-GR}
  r \Psi_{4} \sim \sum_{l m n} \tilde{A}_{l m n} e^{-i\tilde{\omega}_{l mn}(t-r)} e^{im\phi} \, {}_{-2}S_{l m}^{a\tilde{\omega}}(\theta)
  \,.
\end{align}
Here $\tilde{A}_{lmn}$ is a complex mode amplitude with complex frequency
$\tilde{\omega}_{lmn}$.  Here $l,m$ label the angular harmonics, and $n\ge 0$
labels the QNMs with the same $lm$ in terms of increasing damping
time, related to the imaginary part of the complex
frequency (henceforth we suppress the $lmn$ labels for brevity).  The
damping time $\tau > 0$ is found via
\begin{align}
  \label{eq:omega-conv}
  \tilde{\omega} &= \omega - i / \tau \,,
  &
  \tau \equiv -1/\text{Im}[\tilde{\omega}] \,.
\end{align}
The functions $e^{im\phi} \, {}_{-2}S_{lm}^{a\tilde{\omega}}(\theta)$ are spin
$-2$-weighted \emph{spheroidal} harmonics.  However
our numerical simulations compute
$\Psi_{4}^{(0)}$ and $\Psi_{4}^{(2)}$ decomposed into spin-weighted
\emph{spherical} harmonics, where the spheroidals are a deformation of
the sphericals.  The spherical-spheroidal mixing coefficients can be
computed~\cite{Hughes:1999bq, Cook:2014cta, Berti:2015itd, QNMCode},
and the amount of mixing is small at low spins and frequencies, so we
do not model the mixing in this paper.

\subsection{QNMs beyond GR}
\label{sec:QNMs-beyond-gr}

Now suppose we are interested in some (unknown) beyond-GR theory that
is UV complete.  We assume that there is a parameter $\varepsilon$
such that as $\varepsilon\to 0$, this theory recovers GR.  Therefore
we can expand around $\varepsilon=0$ to control the calculation.  If
we keep some leading number of terms, this theory will coincide with
an order-reduced EFT, for example with order-reduced dCS as presented
in Sec.~\ref{sec:dCS}.  There are now two small parameters: the size
of the GW perturbations $\xi$, and the amount of deformation away from
GR, $\varepsilon$.  The metric and any additional new degrees of
freedom will now be a bivariate expansion in $(\varepsilon,\xi)$.  We
will demonstrate this abstractly.

Suppose the nonlinear, UV-complete EOMs can be written as
\begin{align}
  \label{eq:UV-FEs}
  \mathcal{F}[\bs{u}] = 0
  \,,
\end{align}
where $\bs{u}$ is a vector of all the field variables [for example,
any UV completion of dCS must include at least $\bs{u} =
(g_{ab},\vartheta,\ldots)$].  The new degrees of freedom should be
frozen out in the $\varepsilon\to 0$ limit.

We assume these equations have a family of nonlinear solutions for
stationary, axisymmetric BHs
$\bs{u}^{\txt{BH}}(\varepsilon, M, a, \ldots)$.  As in GR, we
conjecture that the final state of a merger of two BHs will be a
unique perturbation to a single BH of this family.  Therefore, we can
perform linear perturbation theory about one solution, positing
\begin{align}
  \label{eq:g-eps-xi}
  g_{ab}(\varepsilon) = g_{ab}^{\txt{BH}}(\varepsilon)
  + \xi h_{ab}(\varepsilon)
  + \mathcal{O}(\xi^{2})
  \,,
\end{align}
and similarly for all other fields in $\bs{u}(\varepsilon) =
\bs{u}^{\txt{BH}}(\varepsilon) + \xi \bs{u}^{(1)}(\varepsilon) +
\mathcal{O}(\xi^{2})$.  Linearizing Eq.~\eqref{eq:UV-FEs} about
$\bs{u}^{\txt{BH}}$ would derive the linear EOMs that
parallel
GR's metric perturbation equations~\eqref{eq:met-pert}, with some
linear PDE
\begin{align}
  \label{eq:UV-FEs-pert}
  \mathcal{F}^{(1)}[\bs{u}^{(1)}] = 0
  \,.
\end{align}
At the level of the ``full'' equations, QNMs are once again
homogeneous solutions to this linear PDE.
Since the background is axisymmetric, the perturbations will again
decompose into azimuthal modes proportional to $e^{im\phi}$, labeled
by $m$.  Since the background is stationary, the perturbations will
also decompose into modes proportional to $e^{-i\tilde{\omega} t}$, labeled by
complex frequencies $\tilde{\omega}$, with $\text{Im}[\tilde{\omega}] < 0$ if we
assume the nonlinear BH solutions are stable.

The QNMs in this beyond-GR theory need not be diagonal in field space
$(g_{ab}, \vartheta, \ldots)$.  For example, mixed QNM modes are
present in dCS, as discussed in~\cite{Molina:2010fb}, and in
Einstein-dilaton-Gauss-Bonnet gravity, as claimed
in~\cite{Blazquez-Salcedo:2016enn, Blazquez-Salcedo:2017txk,
  Pani:2009wy, Witek:2018dmd}
(though see below for further discussion).
However, since all new degrees of
freedom beyond the metric should freeze out in the limit
$\varepsilon\to 0$, the mode structure should become diagonal in field
space in this limit.  Regardless of the diagonal basis, in the
$r\to\infty$ limit, $r\Psi_{4}$ will be a linear combination of the
form
\begin{align}
  \label{eq:rPsi-general}
  r\Psi_{4}(\varepsilon) \sim \sum_{\text{modes }\lambda}
  A_{\lambda}(\varepsilon) e^{-i\tilde{\omega}_{\lambda}(\varepsilon)\,(t-r)}e^{im\phi}
  S_{\lambda}(\varepsilon, \theta) \,,
\end{align}
where $\lambda$ labels the modes.

\subsection{Perturbative treatment of QNMs beyond GR}
\label{sec:pert-treatm-qnms}

We can now ask how these quantities are deformed away from their GR
values.  Notice that in Eq.~\eqref{eq:rPsi-general}, the frequencies
$\tilde{\omega}(\varepsilon)$ and the amplitude of each mode
$\tilde{A}_{\lambda}(\varepsilon)$ should depend continuously on
$\varepsilon$.  Some modes $\lambda$ will reduce to GR modes $lmn$ in
the limit as $\varepsilon\to 0$, and their frequencies and amplitudes
will reduce to the appropriate GR quantities.  Additional modes (e.g.\
the ``scalar-led'' modes of~\cite{Blazquez-Salcedo:2016enn,
  Blazquez-Salcedo:2017txk}) will have amplitudes that vanish at
$\varepsilon=0$.

Now turn to the bivariate expansion of Eq.~\eqref{eq:g-eps-xi},
\begin{align}
  \label{eq:g-bivar}
  \begin{split}
    g_{ab} ={}& (g_{ab}^{K} + \xi h_{ab}^{Q\,(0)}) + \\
    &{}\varepsilon
    (h_{ab}^{\text{Def}} + \xi h_{ab}^{Q\,(1)}) +
    \mathcal{O}(\varepsilon^{2},\xi^{2}) \,.
  \end{split}
\end{align}
Here $h^{\txt{Def}}$ is the deformation of the full BH metric away from
the Kerr metric, $g^{\txt{BH}} = g^{K}+\varepsilon h^{\txt{Def}} +
\mathcal{O}(\varepsilon^{2})$, and similarly the QNMs can be expanded
into their GR parts $h^{Q\,(0)}$ and deformation $h^{Q\,(1)}$.
Doing so requires expanding all quantities in powers of $\varepsilon$,
\begin{align}
  \label{eq:A-pert}
  \tilde{A}_{lmn}(\varepsilon) &= \sum_{k} \varepsilon^{k} \tilde{A}_{lmn}^{(k)} \,, \\
  \label{eq:omega-pert}
  \tilde{\omega}_{lmn}(\varepsilon) &= \sum_{k} \varepsilon^{k} \tilde{\omega}_{lmn}^{(k)} \,,
\end{align}
and similarly for $S_{lmn}(\varepsilon,\theta)$.  The quantities with
superscript ${(0)}$ are the GR expression appearing in
Eq.~\eqref{eq:rPsi-GR}.
The real and imaginary parts of $\tilde{\omega}^{(k)}$ follow the same
expansion, but we also use $\tau \equiv -1/\text{Im}[\tilde{\omega}]$.  Similarly
expanding $\tau = \sum_{k} \varepsilon^{k} \tau^{(k)}$ and using the
chain rule gives
\begin{align}
  \tau^{(1)} = + \frac{\text{Im}[\tilde{\omega}^{(1)}]}{\text{Im}[\tilde{\omega}^{(0)}]^{2}}
  = \text{Im}[\tilde{\omega}^{(1)}] \left(\tau^{(0)}\right)^{2} \,.
\end{align}

If we plug the leading order corrections from Eqs.~\eqref{eq:A-pert}
and \eqref{eq:omega-pert} into the ansatz Eq.~\eqref{eq:rPsi-general},
we will arrive at the form
\begin{widetext}
  \begin{align}
    r\Psi_{4} &= r\Psi_{4}^{(0)} + \varepsilon \, r\Psi_{4}^{(1)} + \ldots \,, \\
    r\Psi_{4}^{(0)} &= \sum_{lmn} 
    \tilde{A}_{lmn}^{(0)}
    e^{-i \tilde{\omega}_{lmn}^{(0)}(t-r)}  e^{im\phi} \, {}_{-2}S_{l m}^{a\tilde{\omega}}(\theta) \,, \\
    \label{eq:rPsi1}
    r\Psi_{4}^{(1)} &= \sum_{lmn} 
    \left(
      \tilde{A}_{lmn}^{(1)} - i \tilde{\omega}_{lmn}^{(1)} (t-r) \tilde{A}_{lmn}^{(0)}
    \right) e^{-i \tilde{\omega}_{lmn}^{(0)}(t-r)}  e^{im\phi} \, {}_{-2}S_{l m}^{a\tilde{\omega}}(\theta)
    + \text{mode mixing term.}
  \end{align}
\end{widetext}
The quantity denoted by ``mode mixing'' is proportional to
$\frac{d}{d\varepsilon}S_{\lambda}(\varepsilon,\theta)$.  Since we are
already ignoring the difference between spherical and spheroidal
harmonics in GR, we also ignore this extra mode mixing term in this
manuscript.

Let us make a few notes about the functional form of $r\Psi_{4}^{(1)}$
in Eq.~\eqref{eq:rPsi1}.  First, although the ``full'' QNM frequency
$\tilde{\omega}(\varepsilon)$ at finite $\varepsilon$ is shifted relative to
$\tilde{\omega}^{(0)}$, everything in Eq.~\eqref{eq:rPsi1} is proportional to
$e^{-i\tilde{\omega}^{(0)}u}$, where  $u \sim t-r$ is retarded time.  Second, Eq.~\eqref{eq:rPsi1} is not just a sum
of damped sinusoids, because of the term going as
$\sim \tilde{\omega}^{(1)} \tilde{A}^{(0)} u e^{-i\tilde{\omega}^{(0)} u}$.  Since $\text{Im}[\tilde{\omega}^{(0)}]<0$, this term remains
bounded, but it grows in importance relative to the term
$\tilde{A}^{(1)} e^{-i\tilde{\omega}^{(0)}u}$.  This is a typical symptom of a
secular breakdown of perturbation theory~\cite{MR538168, MR1392475}.

\subsection{Particular, homogeneous modes, and\\ modes that are mixed in
  field space}
\label{sec:part-homog-modes}

How does the new term $\sim u e^{-i\tilde{\omega}^{(0)}u}$ arise, which
differs in form from both the GR and ``exact'' beyond-GR behaviors?
To understand we have to look at the perturbative treatment of the
equations of motion.
We can model the metric sector of $\mathcal{F}$ as some deformation of
Einstein's equations,
\begin{align}
  \label{eq:G-deform}
  \mpl^{2} G_{ab}[g_{cd}(\varepsilon)] + \varepsilon H_{ab}[\bs{u}] = 0
  \,.
\end{align}
For example in dCS,
$\varepsilon H_{ab} = \mpl \ell^{2 }C_{ab} - T_{ab}^{\vartheta}$ which
is divergence-free on shell.  Inserting Eq.~\eqref{eq:g-bivar} into
Eq.~\eqref{eq:G-deform} and expanding will give order-reduced
equations for the deformations.  For example, setting $\xi=0$ reduces
to the BH background with no QNMs.  The deformation $h^{\txt{Def}}$
satisfies the inhomogeneous equation
\begin{align}
  \mpl^{2} G^{(1)}_{ab}[h_{cd}^{\txt{Def}}] = - H_{ab}[g^{K}_{cd}] \,.
\end{align}
Now including the terms at $\mathcal{O}(\xi)$, we see
\begin{align}
  \label{eq:met-pert-BGR}
  \mpl^{2} (G^{(1)}_{ab}[h_{cd}^{\txt{Def}}]+\xi G^{(1)}_{ab}[h^{Q\,(1)}_{cd}]) =
  - H_{ab}[g^{K}_{cd}+\xi h^{Q\,(0)}_{cd}]
  \,.
\end{align}
At linear order in $\xi$, $H_{ab}^{(1)}[h^{Q\,(0)}_{cd}]$ (expanded about
$g^{K}$) will generate a source term for $h^{Q\,(1)}$.  In our
numerical implementation we 
have the full GR metric solution, not
an expansion in powers of $\xi$, so there can also be nonlinear
(QNM)$^{2}$ and higher terms appearing in the source.

This equation is different from the linearized EOM in
GR~\eqref{eq:met-pert} or in the ``exact''
theory~\eqref{eq:UV-FEs-pert}: those are both homogeneous, whereas
Eq.~\eqref{eq:met-pert-BGR} is inhomogeneous (it has a source term on
the RHS).  The solution will therefore have two parts: a homogeneous
solution, and a particular solution.  The linear differential operator
$G^{(1)}_{ab}[h_{cd}^{Q\,(1)}]$ appearing on the LHS of
Eq.~\eqref{eq:met-pert-BGR} is the same operator, expanded about the
same Kerr background, as in the GR Eq.~\eqref{eq:met-pert}.  Therefore
the homogeneous solutions are the same.  This accounts for the
$e^{-i\tilde{\omega}^{(0)}u}$ terms in Eq.~\eqref{eq:rPsi1}.  The
$u e^{-i\tilde{\omega}^{(0)}u}$ terms are particular solutions.  Their secular
behavior is due to the source term having support at the poles of
the Green's function. 

In~\cite{Witek:2018dmd}, the authors simulated the leading-order
scalar field behavior on a binary black hole background in
Einstein-dilaton-Gauss-Bonnet gravity, and found that the solution for
the scalar field during ringdown contained two parts, similar to how
we observed two pieces in Sec.~\ref{sec:observe-part-homog-solut}.
The authors referred to these as ``scalar-led'' and
``gravitational-led'' modes which they suggest are due to mixing in
field space.

This nomenclature was introduced in~\cite{Blazquez-Salcedo:2016enn}
but was seen earlier in dCS in Molina et al.~\cite{Molina:2010fb}.
In~\cite{Molina:2010fb}, the authors investigated QNMs of
Schwarzschild black holes in \textit{full} dCS gravity. For zero spin, the
system is well-posed, and thus can be solved in the full theory,
without working in an order-reduction or other perturbative
scheme.\footnote{%
  Working on a Schwarzschild background within the order-reduction
  scheme also faithfully reproduces $\vartheta^{(1)} = 0$ and
  $g_{ab}^{(2)} = 0$.}
The radial parts of the scalar and gravitational QNMs for each mode
are governed by a set of \textit{fully coupled} ordinary differential equations (ODEs) of the form
(cf. Eqs.~(2.8) and (2.9) in~\cite{Molina:2010fb}),
\begin{align}
  \label{eq:coupled-vartheta-Psi}
  \frac{d^2}{dr_*^2} \begin{pmatrix} 
\vartheta \\
\Psi
\end{pmatrix} = 
\begin{pmatrix}
V_{11} & V_{12} \\ V_{21} & V_{22}
\end{pmatrix}
\begin{pmatrix} 
\vartheta \\
\Psi
\end{pmatrix}\,,
\end{align}
where $r_*$ is the usual Schwarzschild ``tortoise'' radial coordinate,
$\vartheta$ is a spherical harmonic mode of the full dCS scalar field,
$\Psi$ is a spherical harmonic mode of the metric, and the $V_{ij}$
potentials are functions dependent on $r$ and the dCS coupling
parameter.  Since $\vartheta$ and $\Psi$ are coupled in this
2-dimensional linear ODE problem,  we find two types of QNMs
(scalar-led and gravitational-led, which become diagonal in field
space in the limit $\ell\to 0$).

However, we and~\cite{Witek:2018dmd} are both working in a
perturbative scheme.  In the perturbation scheme, the leading-order
dCS metric perturbation $h_{ab}^{(2)}$ does not back-react onto the
leading scalar field $\vartheta^{(1)}$.  If one applies the
perturbation scheme to Eq.~\eqref{eq:coupled-vartheta-Psi} on the
Schwarzschild background, the ODEs take the form
\begin{align}
  \frac{d^2}{dr_*^2} \begin{pmatrix} 
\vartheta^{(1)} \\
\Psi^{(2)}
\end{pmatrix} =
\begin{pmatrix}
W_{11} & 0 \\ W_{21} & W_{22}
\end{pmatrix}
\begin{pmatrix} 
\vartheta^{(1)} \\
\Psi^{(2)}
\end{pmatrix}\,.
\end{align}
This matrix is triangular, so the solution for $\vartheta^{(1)}$ can
be found independently of $\Psi^{(2)}$.  Meanwhile, the QNMs of
$\vartheta^{(1)}$ enter into the source term for $\Psi^{(2)}$.

Thus, the presence of the ``scalar-led'' mode (now on a nonlinear,
perturbed Kerr background) seen by~\cite{Witek:2018dmd} is not
surprising, because the homogeneous solution of the beyond-GR scalar
perturbation equation will contain the same scalar QNMs (homogeneous
solutions) as in GR.  Meanwhile, the ``gravitational-led'' mode in
$\vartheta^{(1)}$ is surprising.  However, looking back at
Eq.~\eqref{eq:met-pert-BGR} suggests its origin (keeping in mind that
everything in order-reduced dCS and EDGB is pushed up by a power of
$\varepsilon$, so we should think of Eq.~\eqref{eq:met-pert-BGR} of
suggesting the form of the scalar equation for $\vartheta^{(1)}$).
The nonlinear, perturbed Kerr background enters the source term on the
RHS of \eqref{eq:met-pert-BGR}, generating a source that oscillates at
the frequency of the background (GR) gravitational waves.  This
sources a \emph{particular} solution at this frequency.

Therefore we conjecture that the ``gravitational-led'' mode appearing
the scalar field in~\cite{Witek:2018dmd} was actually a particular
solution, though it is fair to still consider it as part of the QNM
spectrum.  We can more generally conjecture that when the perturbative
approach is applied to field-space mixed QNM modes, they will appear
as combinations of homogeneous and particular solutions of the linear
equations.  Investigating this conjecture is beyond the scope of this
work.


\section{Choosing a perturbed gauge}
\label{sec:GaugeAppendix}

Throughout this appendix, as well as Appendix~\ref{sec:WaveExtractionAppendix}, we use the notation developed in~\cite{MashaEvPaper}, and standard 3+1 ADM decomposition notation~\cite{baumgarteShapiroBook}. Recall that $g_{ab}$ refers to the 4-dimensional spacetime metric, while $\gamma_{ij}$ refers to the 3-dimensional spatial metric. $\Delta Q$ is the leading-order perturbation to quantity $Q$.

The generalized harmonic evolution for the background follows from the equation
\begin{align}
\label{eq:GH}
\Gamma_a = -H_a\,,
\end{align}
where $\Gamma_a \equiv g^{bc} \Gamma_{bca}$, and $H_a$ is known as the \textit{gauge source function} (cf.~\cite{Lindblom2006} for more details). Throughout the evolution, the \textit{gauge constraint},
\begin{align}
    \label{eq:GaugeConstraint}
C_a \equiv H_a + \Gamma_a = 0\,,
\end{align}
must be satisfied. 

When generating initial data for $g_{ab}$ and $\pd_t g_{ab}$, we are free to choose $\pd_t \alpha$ and $\pd_t \beta^i$, the initial time derivatives of the lapse and shift. These quantities appear in $\Gamma_a$, so choosing them is equivalent to choosing initial values of $H_a$, via Eq.~\eqref{eq:GaugeConstraint}. For example, for initial data in equilibrium, we can set $\pd_t \alpha = 0$ and $\pd_t \beta^i  = 0$, and set $H_a$ to initially satisfy Eq.~\eqref{eq:GaugeConstraint}. Alternatively, we can choose to work in a certain gauge, such as harmonic gauge with $H_a = 0$, and set $\pd_t \alpha$ and $\pd_t \beta^i$ to satisfy Eq.~\eqref{eq:GaugeConstraint}.  

As the evolution progresses, we can either leave $H_a$ fixed, or continuously `roll' it into  a  different gauge, with the restriction that it contains only up to first derivatives of $g_{ab}$ to ensure well-posedness. In practice, for BBH in GR, we work in a \textit{damped harmonic gauge}, with $H_a$ specified using the methods given in~\cite{Szilagyi:2009qz}.

The perturbed generalized harmonic evolution takes a similar form as Eq.~\eqref{eq:GH}, with
\begin{align}
    \Delta \Gamma_a = - \Delta H_a \,,
\end{align}
where $\Delta \Gamma_a$ is the first-order perturbation to $\Gamma_a$, and $\Delta H_a$ is a perturbed gauge source function. Similar to Eq.~\eqref{eq:GaugeConstraint}, we have a perturbed gauge constraint,
\begin{align}
\label{eq:PerturbedGaugeConstraint}
    \Delta C_a \equiv \Delta \Gamma_a + \Delta H_a = 0\,.
\end{align}
At the start of the evolution, we similarly have the freedom to choose $\Delta H_a$, provided that it contains no higher than first derivatives of $\Delta g_{ab}$, and satisfies the perturbed gauge constraint Eq.~\eqref{eq:PerturbedGaugeConstraint}. When solving for perturbed initial data (cf.~\cite{MashaIDPaper}), we similarly have the freedom to choose $\pd_t \Delta \alpha$ and $\pd_t \Delta \beta^i$, the time derivatives of the perturbed lapse and shift. An easy choice, for example, is to work in a perturbed harmonic gauge, \begin{align}
    \label{eq:PerturbedHarmonic}
    \Delta H_a = 0\,.
\end{align}

Let us now work out how to set $\pd_t \Delta \alpha$ and $\pd_t \Delta \beta^i$ in order to satisfy Eq.~\eqref{eq:PerturbedGaugeConstraint} for some desired perturbed gauge source function $\Delta H_a$. Let us first consider the unperturbed case, setting $\pd_t \alpha$ and $\pd_t \beta^i$ for some gauge source function $H_a$. We will work with the $\kappa_{abc}$ variable, which is the fundamental variable encoding the spatial and time derivatives of the metric (cf.~\cite{MashaEvPaper}) as
\begin{align}
    \kappa_{iab} &\equiv \pd_i g_{ab}\,,  \\
    \kappa_{0ab} &\equiv -n^c \pd_c g_{ab}\,,
\end{align}
where $n^c$ denotes the timelike unit normal vector. We can use our freedom to set $\pd_t \beta^i$ and $\pd_t \alpha$ to modify $\kappa_{abc}$ to satisfy $\Gamma_a = -H_a$ as 
\begin{align}
\label{eq:kappa00i}
\kappa_{00i} = -\alpha H_i + \beta^k \kappa_{0ki} - \alpha \gamma^{jk} \Gamma_{ijk} - \frac{1}{2} \alpha n^a n^b \kappa_{iab}\,,
\end{align}
where $\Gamma_{ijk}$ is the spatial Christoffel symbol of the first kind, and
\begin{align}
\label{eq:kappa000}
    \kappa_{000} &= -2 \alpha H_0 + 2 \beta^j (\kappa_{00j} + \alpha H_j) \\
    \nn & \quad - \beta^j \beta^k \kappa_{0jk} -  \alpha^2 \gamma^{jk} \kappa_{0jk} - 2 \alpha^2 \gamma^{jk} n^a \kappa_{jka} \,,
\end{align}
where $\kappa_{00j}$ in the above expression is given by Eq.~\eqref{eq:kappa00i}. We can then use this modified $\kappa_{abc}$ to compute $\Gamma_a$ and ensure that Eq.~\eqref{eq:GaugeConstraint} is satisfied for $H_a = H_a$.

Perturbing Eqs.~\eqref{eq:kappa00i} and~\eqref{eq:kappa000}, we can get an expression for a modified $\Delta \kappa_{abc}$ to satisfy Eq.~\eqref{eq:PerturbedGaugeConstraint} for some desired perturbed gauge source function $\Delta H_a$. We thus obtain
\begin{align}
\label{eq:hkappa00i}
\Delta \kappa_{00i} &= -\Delta \alpha H_i -\alpha \Delta H_i \\
\nn & \quad + \Delta \beta^k \kappa_{0ki} + \beta^k \Delta \kappa_{0ki} \\
\nn & \quad - \Delta \alpha \gamma^{jk} \Gamma_{ijk} - \alpha \Delta \gamma^{jk} \Gamma_{ijk} - \alpha \gamma^{jk} \Delta \Gamma_{ijk} \\
\nn & \quad - \frac{1}{2} \Delta \alpha n^a n^b \kappa_{iab} - \frac{1}{2} \alpha \Delta n^a n^b \kappa_{iab} \\
\nn & \quad - \frac{1}{2} \alpha n^a \Delta n^b \kappa_{iab} - \frac{1}{2} \alpha n^a n^b \Delta \kappa_{iab}\,,
\end{align}
and 
\begin{align}
\label{eq:hkappa000}
    \Delta \kappa_{000} &= -2 \Delta \alpha H_0 -2 \alpha \Delta H_0 \\
    \nn & \quad + 2 \Delta \beta^j (\kappa_{00j} + \alpha H_j) \\ 
    \nn & \quad + 2 \beta^j (\Delta \kappa_{00j} + \Delta \alpha H_j + \alpha \Delta H_j) \\
    \nn & \quad - \Delta \beta^j \beta^k \kappa_{0jk} - \beta^j \Delta \beta^k \kappa_{0jk}  - \beta^j \beta^k \Delta \kappa_{0jk}  \\
    \nn & \quad -  2 \alpha \Delta \alpha \gamma^{jk} \kappa_{0jk} -  \alpha^2 \Delta \gamma^{jk} \kappa_{0jk} -  \alpha^2 \gamma^{jk} \Delta \kappa_{0jk} \\
    \nn & \quad - 4 \alpha \Delta \alpha \gamma^{jk} n^a \kappa_{jka} - 2 \alpha^2 \Delta \gamma^{jk} n^a \kappa_{jka} \\
    \nn & \quad - 2 \alpha^2 \gamma^{jk} \Delta n^a \kappa_{jka} - 2 \alpha^2 \gamma^{jk} n^a \Delta \kappa_{jka}\,.
\end{align}
Note that this computation also uses the gauge source function of the background, $H_a$. Assuming that the background is in a satisfactory gauge, we set $H_a$ to the initial background gauge source function. All of the perturbed quantities in Eqs.~\eqref{eq:hkappa00i} and~\eqref{eq:hkappa000} are given in~\cite{MashaEvPaper}.

In this study, we choose to work in a perturbed harmonic gauge, with $\Delta H_a = 0$.


\section{Computing perturbed gravitational radiation}
\label{sec:WaveExtractionAppendix}

The outgoing gravitational radiation of a  spacetime is encoded in the Newman-Penrose scalar $\Psi_4$. In order to compute the leading-order correction to the binary black hole background radiation due to the metric perturbation $\Delta g_{ab}$, we need to compute $\Delta \Psi_4$, the leading-order correction to $\Psi_4$. 

$\Psi_4$, a scalar, is computed on a topologically spherical surface from a rank-two tensor $U_{ij}$, contracted with a tetrad (in our case, a coordinate tetrad that converges to a quasi-Kinnersley tetrad at large radii). $U_{ij}$ on a surface with normal vector $\hat{n}^i$ takes the form
\begin{align}
\label{eq:UEB}
U_{ij} &= (P_i^m P_j^n -\frac{1}{2}P_{ij}P^{mn})(E_{mn} - \epsilon_m{}^{kl}\hat{n}_l B_{kn}) \,,
\end{align}
where $E_{ij}$ is the electric Weyl tensor, $B_{ij}$  is the magnetic Weyl tensor,  $\epsilon_{ijk}$ is the (spatial) Levi-Civita  tensor, and  the projection  operators  are  given  by
\begin{align}
P^{ij} &= \gamma^{ij} - \hat n^i \hat n^j \,, \\
P_{ij} &= \gamma_{ij} - \hat n_i \hat n_j \,, \\
P^i_j &= \gamma^i_j - \hat n^i \hat n_j \,.
\end{align}
Here, the vector $\hat n^i$ and the one form $\hat n_i$ are normalized using $N \equiv \sqrt{\gamma^{ij} n_i n_j}$ with $n^i = \gamma^{ij}n_j$. 

In order to perturb $\Psi_4$, let us write the electric and magnetic Weyl tensors in Eq.~\eqref{eq:UEB} in terms of the extrinsic curvature $K_{ij}$,
\begin{align}
\label{eq:URK}
U_{ij} &= (P_i^m P_j^n -\frac{1}{2}P_{ij}P^{mn}) \times \\
\nn & \quad \Big(R_{mn} + \gamma^{kl}(K_{mn}K_{kl} - K_{mk}K_{ln}) \\
\nn & \quad  - \hat n^k (D_k K_{mn} - D_{(m} K_{n) k} )\Big)\,,
\end{align}
where $R_{ij}$ is the spatial Ricci tensor and $D_i$ is the spatial covariant derivative associated with $\gamma_{ij}$. 

Perturbing Eq.~\eqref{eq:URK}, we obtain
\begin{align}
    \Delta U_{ij} &= (P_i^m P_j^n -\frac{1}{2}P_{ij}P^{mn}) \times \\
    \nn & \quad \Big (\Delta R_{mn} + \Delta \gamma^{kl}(K_{mn}K_{kl} - K_{mk}K_{ln}) \\
    \nn & \quad + \gamma^{kl}(\Delta K_{mn}K_{kl} + K_{mn} \Delta K_{kl} \\
    \nn & \quad  - \Delta K_{mk}K_{ln} - K_{mk}\Delta K_{ln})\\
\nn &  \quad - \hat n^k (\Delta(D_k K_{mn}) - \Delta(D_{(m} K_{n) k} ) \\
\nn & \quad - \Delta \hat n^k (D_k K_{mn} - D_{(m} K_{n) k} )\Big) \\
\nn & \quad + (\Delta P_i^m P_j^n + P_i^m \Delta P_j^n \\ 
\nn & \quad -\frac{1}{2}\Delta P_{ij}P^{mn} -\frac{1}{2}P_{ij}\Delta P^{mn}) \times U_{mn}
\,.
\end{align}
All of the perturbed quantities $\Delta g^{ij}, \Delta K_{ij}, \Delta (D_k K_{ij})$, and $\Delta R_{ij}$ are given in terms of the perturbation to the spatial metric, $\Delta \gamma_{ij} = \Delta g_{ij}$,  its spatial derivative $\pd_k \Delta \gamma_{ij} = \pd_k \Delta g_{ab}$, and its time derivative, $\pd_t \Delta \gamma_{ij} = \pd_t \Delta g_{ij}$ in~\cite{MashaIDPaper}. Note that since we use a first-order scheme, we have access to $\Delta g_{ab}$, $\pd_c \Delta g_{ab}$ throughout the evolution (cf.~\cite{MashaEvPaper}). 

Let us now work through the perturbations to the normal vectors and projection operators. Because we want the perturbation to the gravitational radiation to be extracted on the same surface as the background gravitational radiation, we will hold the unnormalized one-form to the surface, $n_i$, fixed. That is, $\Delta n_i = 0$. From this, we can then compute
\begin{align}
    \Delta N &= \Delta (\gamma^{ij} n_i n_j)^{1/2} = \frac{1}{2} \Delta \gamma^{ij} n_i n_j (\gamma^{ij} n_i n_j)^{-1/2} \\
    \nn &= \frac{1}{2 N} \Delta \gamma^{ij} n_i n_j \,,
\end{align}
and 
\begin{align}
    \Delta \hat{n}_i &= - \frac{n_i}{N^2} \Delta N = -\frac{\hat{n}_i}{N} \Delta N\,, \\
    \Delta n^i &= \Delta \gamma^{ij} n_j = \Delta \gamma^{ij} \hat{n}_j N\,,  \\
    \Delta \hat{n}^i &= \frac{\Delta n^i}{N} - \frac{n^i}{N^2} \Delta N \\
    \nn &= \frac{\Delta \gamma^{ij} n_j}{N} - \frac{n^i}{N^2} \Delta N \\
    \nn &= \Delta \gamma^{ij} \hat{n}_j - \frac{\hat{n}^i}{N} \Delta N  \,.
\end{align}
We can then perturb the projection operators,
\begin{align}
\Delta P^{ij} &= \Delta \gamma^{ij} - \Delta  \hat n^i \hat n^j - \hat n^i \Delta \hat n^j \,, \\
\Delta P_{ij} &= \Delta \gamma_{ij} - \Delta \hat n_i \hat n_j  - \hat n_i \Delta \hat n_j \,, \\
\Delta P^i_j &= \Delta \gamma^i_j - \Delta \hat n^i \hat n_j  - \hat n^i \Delta \hat n_j \,,
\end{align}
where $\Delta \gamma^i{}_j = \Delta \gamma^{ik} \gamma_{kj} + \gamma^{ik} \Delta \gamma_{kj} $

Once we obtain $\Delta U_{mn}$, we use the same tetrad to generate $\Delta \Psi_4$ from $\Delta U_{ij}$ as we do for $\Psi_4$.


\bibliography{dCS_paper}
\end{document}